\documentclass[12pt]{article}

\usepackage{amssymb}
\usepackage[centertags]{amsmath}
\usepackage{rotating}
\usepackage[dvips]{color}	

\usepackage{epsf}
\usepackage{pazh_eng}

\tightenlines

\DeclareMathOperator{\mex}{\mathsf M} 
\DeclareMathOperator{\var}{\mathsf D} 
\newcommand{\me}{\mathop{\sf Me}\nolimits} 

\sloppypar

\begin{document}

{\renewcommand{\thefootnote}{\fnsymbol{footnote}}

\title{\bf Numerical Study of Statistical Properties of the Galactic Center Distance
Estimate from the Geometry of Spiral Arm Segments}

\author{\bf
I. I. Nikiforov\footnote{E-mail: \tt nii@astro.spbu.ru} and A. V. Veselova}

}
\setcounter{footnote}{0}

\vspace{-0.9em}
\begin{center}
{\it St.\ Petersburg State University, Universitetskii pr.\ 28,
\\

 Staryi Peterhof, St.\ Petersburg, 198504 Russia}
\end{center}

\vspace{-0.3em}

\sloppypar 
\vspace{2mm}
\noindent
{\bf Abstract}---The influence of various factors on the statistical properties of the Galactic center distance ($R_0$)
estimate obtained by solving the general problem of determining the geometric parameters of a Galactic
spiral arm from its segment with the inclusion of the distance to the spiral pole, i.e., $R_0$, in the set of
parameters has been studied by the Monte Carlo method. Our numerical simulations have been performed
for the model segments representing the Perseus and Scutum arms based on masers in high-mass star
forming regions. We show that the uncertainty in the present-day parallax measurements for these objects
systematically decreases (!) with increasing heliocentric distance, while the relative uncertainty in the
parallaxes is, on average, approximately constant. This lucky circumstance increases significantly (by a
factor of $1.4$--$1.7$) the accuracy of estimating $R_0$ from the arm segment traced by masers. Our numerical
experiments provide evidence for the consistency of the $R_0$ estimate from the spiral-segment geometry.
The significant biases of the estimate detected only for the Scutum arm are caused mainly by the random
parallax errors, the small angular extent of the segment, and the small number of objects representing it.
The dispersion of the $R_0$ estimate depends most strongly on the angular extent of the segment and the
parallax uncertainty if the latter, on average, does not depend on the distance. The remaining parameters,
except for the pitch angle, exert an equally significant, but weaker influence on the statistical accuracy of the
estimate. When the data on $3$--$8$ segments are processed simultaneously, the predicted standard error of the
final estimate is $\sigma_{R_0} \simeq 0.5$--$0.3$~kpc, respectively. The accuracy can be improved by increasing the extent of
the identified segments and the number of objects belonging to them. The method of determining $R_0$ from
spiral segments has turned out to be operable for a wide set of possible parameters even when using an
L-estimator (median). This makes the development of a more complex method based on an M-estimator,
which allows one to properly take into account the measuring and natural dispersions of objects relative to
the arm center line and, thus, to avoid the biases of the parameter estimates, meaningful.


\clearpage

\section{INTRODUCTION}
\label{introduction}

Recent progress in the observational data has
made it possible to distinguish features in the Galactic spiral structure that are {\em{sharp}\/} in the sense of a
strong apparent concentration to their center lines
and simultaneously {\em{extended}\/} (for more details, see
Nikiforov and Veselova 2018; hereafter NV18). This,
in turn, has stimulated works on the spatial modeling
of {\em{separate}\/} spiral arm segments when abandoning
traditional assumptions for studies of the Galactic
spiral pattern, about the number of arms in the Galaxy
and the equality of the pitch angles for different
arms (see NV18 for a review as well as, e.g., the
recent papers by Griv et al.\ (2017) and Krishnan
et al.\ (2017) and references therein). In our previous
paper (NV18) we proposed to abandon yet another
traditional assumption---the preadopted distance $R_0$
from the Sun to the pole of the spiral arms, i.e., the
distance to the Galactic center. If in this problem $R_0$
is deemed a free parameter, along with the segment
parameters, then this, on the one hand, allows the
spiral segments to be modeled in a more general form,
which takes into account, in particular, the noticeable
correlation between $R_0$ and the pitch angle (shown
in NV18), and, on the other hand, potentially opens a
new approach to determining $R_0$ in the class of spatial
methods (for a review, see Nikiforov 2004)---from the
geometry of spiral arms.

A simplified (three-point) method of solving the
problem was developed in NV18 to test this approach.
Its application in the same paper to a sample of objects (masers with trigonometric parallaxes), on the
whole, confirmed that the new approach is operable:
from masers in two arms (Perseus and Scutum) we
managed to obtain reliable results based on which we
deduced an estimate of $R_0 = 8.8 \pm 0.5$~kpc.

In this paper we use the three-point method to
numerically study the statistical properties of the $R_0$
estimate from the geometry of a spiral arm as a function of problem parameters. The method consists in
obtaining the solution for three segment parameters,
including $R_0$, from the positions of its three representative points. Being relatively easy to implement, this
simplified method allows a large number of numerical
experiments to be performed in a reasonable time.
The latter is important for achieving the main goal
of this paper---to evaluate the applicability conditions
and the capabilities of the new approach to finding $R_0$
both for present-day data and in prospect.

\section{MONTE CARLO SIMULATIONS OF THE
$R_0$ MEASUREMENT \\ FROM THE SPIRAL-ARM GEOMETRY}
\subsection{Model Segment Parameters}

In accordance with the parametrization introduced in NV18, we will describe the spiral arm
segment by the following set of parameters: the distance from the coordinate origin, i.e.,
from the Sun, to the pole of the model logarithmic spiral ($R_0$); its pitch angle~($i$);
the spiral position angle at the solar circle $R = R_0 (\lambda_0)$, where $R$ is the
distance from a point in the Galactic plane to the spiral pole; the boundaries of the
spiral segment in Galactocentric longitude ($\lambda^s_1$ and $\lambda^s_2$) that define
the angular extent of the segment ($\Delta \lambda  = \lambda^s_2 - \lambda^s_1$); the
number of objects representing the segment ($	N$); the natural root-mean-square (rms)
scatter of objects across the segment ($\sigma_{\text{w}}$); the absolute
($\sigma_{\varpi}$) or relative ($\sigma_{\varpi}/\varpi$) rms measurement error of the
parallax~$\varpi$. Here the longitudes $\lambda$ are measured in the Galactic plane from
the sunward direction clockwise when viewed from the North Galactic Pole, i.e., in the
direction of Galactic rotation.

The introduction of errors into the parallaxes
as allowance for the uncertainty in the~heliocentric
distances $r$ corresponds here to the use of data
on masers with trigonometric parallaxes, the most
promising type of reference objects for applying the
new method (NV18). The variant with the specification of σ corresponds to the assumption that the
mean parallax uncertainty within the segment may
be deemed approximately constant. However, our
analysis of the data from Reid et al.\ (2014) (see Subsection~2.2) showed that the absolute uncertainty~$\sigma_{\varpi}$
of the present-day parallax measurements for masers
depends systematically on their distance (decreases
with increasing $r$!), i.e., it is generally not constant,
on average, for the segment. On the other hand, we
found that the relative parallax uncertainty changes
more weakly within the limits of spiral segments
revealed by masers. Therefore, the variant with the
specification of~$\sigma_{\varpi}/\varpi$ describes better the present-day data for these objects. Since this tendency for
the parallax error may not be retained in future, we
considered both these variants of parameterizing the
distance uncertainty.

The immediate goal of our numerical simulations
is to find the mean bias and dispersion of the $R_0$
estimate from the spiral-segment geometry as a function of significant problem parameters  ($\Delta\lambda$,
$\sigma_{\text{w}}$, $\sigma_{\varpi}$, $\sigma_{\varpi}/\varpi$, $N$, $i$).

\subsection{Uncertainty in the Trigonometric Parallaxes from Masers}

To choose the representations of the parallax measurement errors in our numerical
experiments, we analyzed the data from two catalogues containing estimates of the
parallaxes~${\varpi}$ and their uncertainties~$\sigma_{\varpi}$ for objects including
masers: the more homogeneous catalogue by Reid et al.\ (2014), which contains 103
high-mass star forming regions (\mbox{HMSFRs}), and the catalogue by Rastorguev et al.\
(2017, hereafter the Rs17 catalogue), which includes the catalogue by Reid et al.\ (2014)
with the replacement of old data by new ones for two objects (hereafter the Rd14
catalogue), as well as the data on 38 additional maser sources, out of which (as our study
of the literature showed) only 9 belong to the class of HMSFRs, while the remaining ones
belong to 14 other types (a total of 141 objects). We use the Rd14 catalogue, because Reid
et al.\ (2014) pointed out the membership of masers in a particular spiral arm, and to
compare the characteristics of the Rd14 objects and the sources added in Rs17.

Table~1 gives the median absolute ($\me \sigma_{\varpi}$)  and
relative ($\me (\sigma_{\varpi}/\varpi)$) parallax uncertainties found by
us separately for the arm segments that were identified by Reid et al.\ (2014) for the entire Rd14 sample
and for the HMSFR and non-HMSFR masers of the
Rs17 catalogue. Since we will also need to choose~$\sigma_{\text{w}}$\,,
the NV18 estimates of the natural scatter of masers
across the segments for the arms for which we managed to resolve this parameter are listed in Table~1.

The $\sigma_{\varpi}=\text{const}$  model for a segment corresponds
to the experience of stellar parallax measurements.
For example, the median of the standard parallax
errors in the Hipparcos catalogue is virtually independent of the stellar magnitude and begins to increase only as the limiting magnitude of the catalogue
is approached (Fig.~1 in Mignard~2000), implying
that the mean parallax uncertainty is approximately
constant as one recedes from the Sun and increases
only at distances close to the limiting ones for the
catalogue. However, for masers the behavior of the
mean~$\sigma_{\varpi}$ with distance turned out to be much better
than the expected one by analogy with stars. $\me\sigma_{\varpi}$
in Table~1 decreases as the segment recedes from
the Sun, taking the greatest value for the Local arm
and the smallest ones for the Scutum arm (in the
inner Galaxy) and the Outer arm (outside the solar
circle). A direct comparison of the catalogued~$\sigma_{\varpi}$ and
$r=1/\varpi$ (Figs.~1a and~2a) shows that the parallax
uncertainty, on average, drops (!) with distance (approximately hyperbolically), with the mean~$\sigma_{\varpi}$ levels
at the boundaries of the~$r$ interval represented by the
data differing by several times for both catalogues.

\begin{figure}
\vspace{-2.1em}

\centerline{%
\epsfxsize=16.0cm%
\epsffile{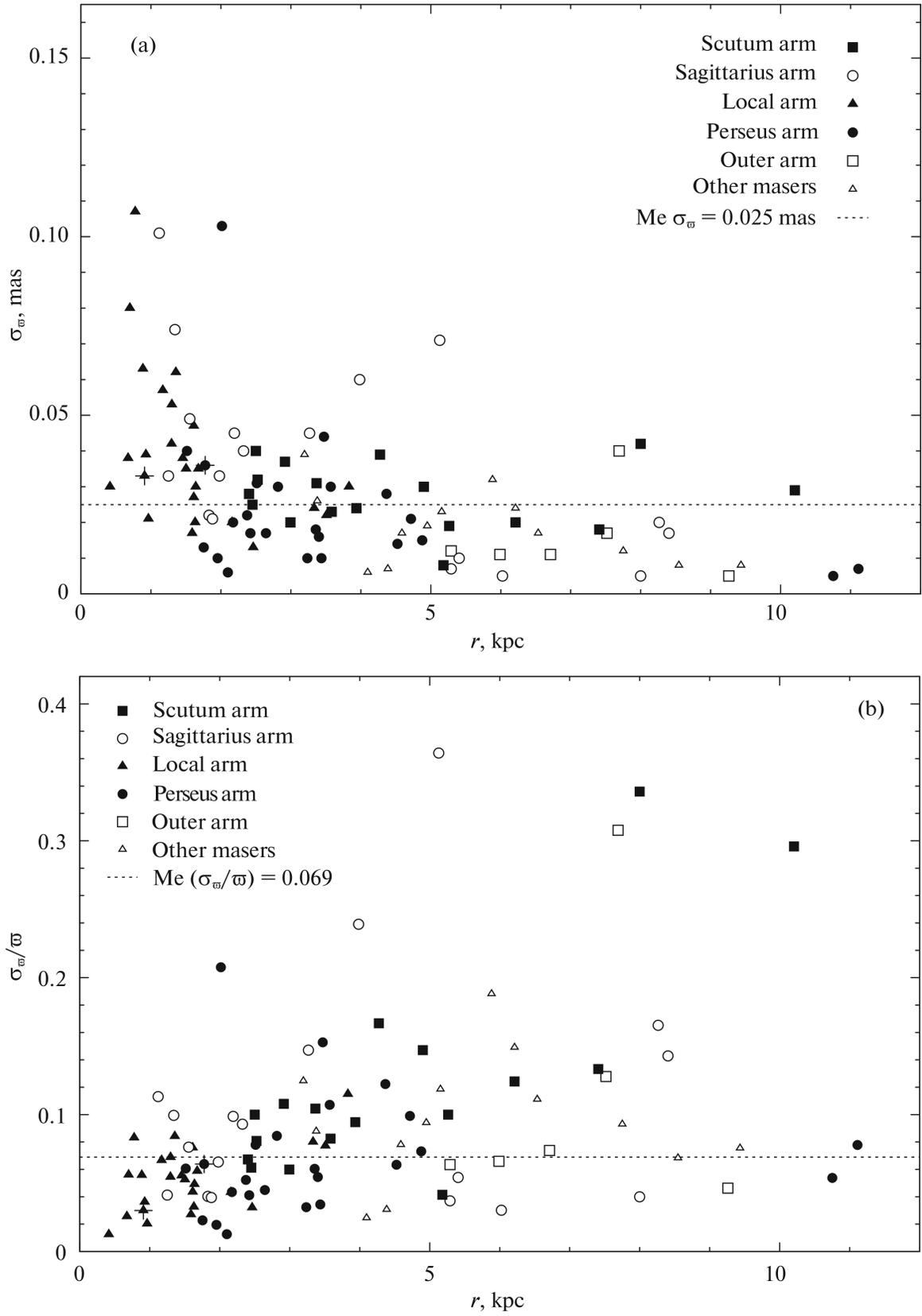}%
}%

\caption{\rm Absolute (a) and relative (b) uncertainties of the parallax measurements versus heliocentric distance for the Rd14
masers. The crosses mark the two objects the data on which were replaced with the newer ones from the Rs17 catalogue. The
dashed lines mark the median values of 
$\sigma_{\varpi}$ and
$\sigma_{\varpi}/\varpi$ for the complete sample of masers in the catalogue.}
\label{spp_sp_r_Rd}
\end{figure}

\begin{figure}
\vspace{-2.1em}

\centerline{%
\epsfxsize=14.8cm%
\epsffile{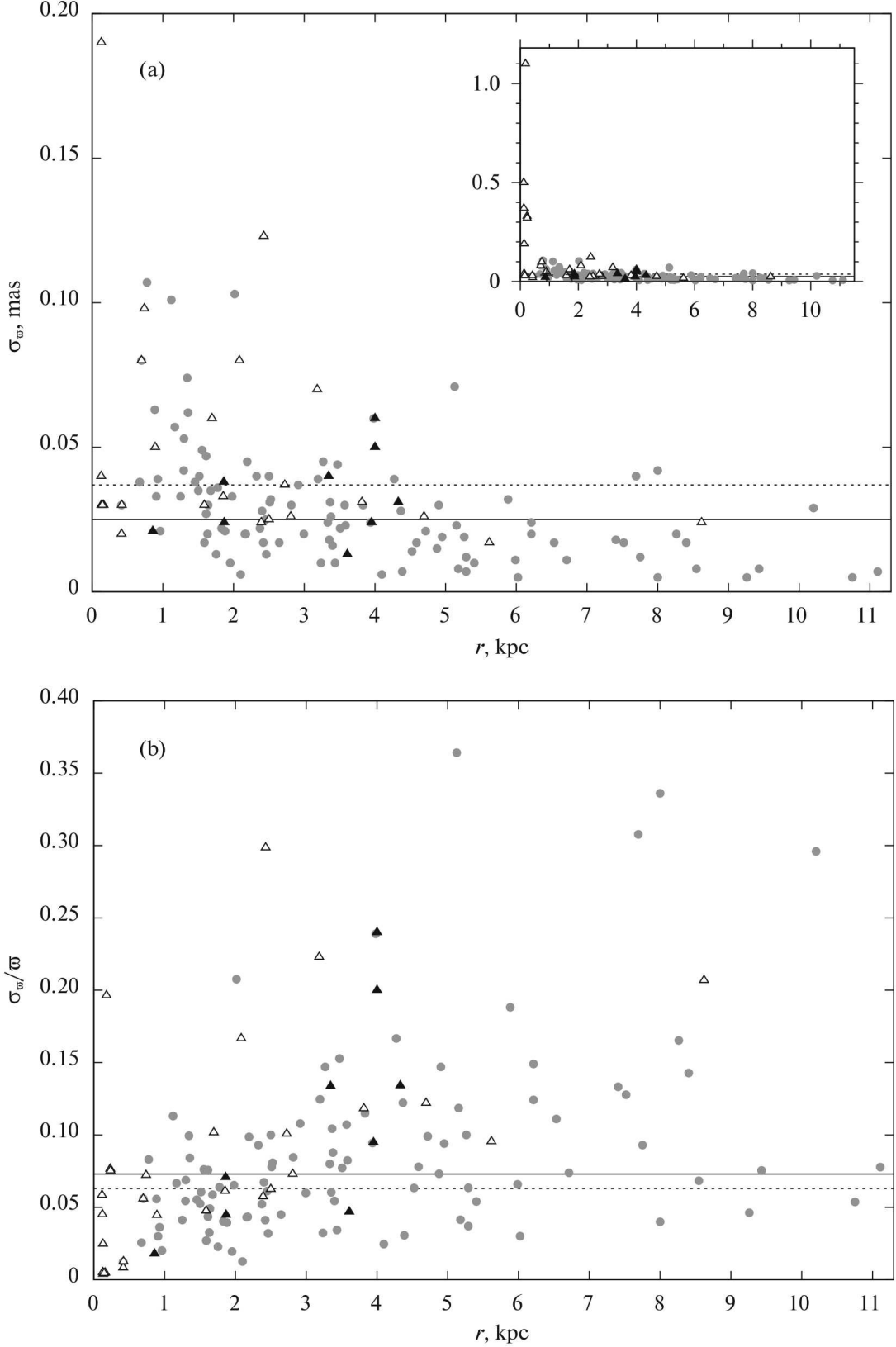}%
}%

\caption{\rm Absolute (a) and relative (b) uncertainties of the parallax measurements versus heliocentric distance for the Rd14
masers (gray circles) and the additional Rs17 masers, HMSFRs (black triangles) and the remaining sources (white triangles).
The solid and dashed lines mark the median values of $\sigma_{\varpi}$ and  $\sigma_{\varpi}/\varpi$  for the complete sample of HMSFR masers in the Rs17 catalogue and for the remaining objects of the same catalogue, respectively.}
\label{spp_sp_r_Rs}
\end{figure}

\begin{table}[t]
 \centering
 \small
 \tabcolsep0.4cm
 \caption{\label{segm_par} \rm Dispersion characteristics of the spatial positions of masers from the data of \newline   the catalogues by Reid et al.\ (2014) and Rastorguev et al.\ (2017)
}
 \vspace{5 pt}
 \begin{tabular}{l|c|c|c|c}
 \hline
 Arm/sample, catalogue & $N$  & $\me \sigma_{\varpi}$,~mas  & $\me (\sigma_{\varpi}/\varpi)$   & $\sigma_{\text{w}}$,~kpc\\[0.05cm]
 \hline
 Scutum  (Sct)\color{black}, Rd14 & $17$  & $0.028$ & $0.100$ &     \\[0.2cm]
 Sagittarius  (Sgr)\color{black}, Rd14 & $18$ &  $0.033$ & $0.085$   & $0.29\pm0.08$\\[0.2cm] 
 Local   (Loc)\color{black}, Rd14 & $25$ &   $0.035$ & $0.05\color{black}4$  & $0.30\pm0.05$  \\[0.2cm]
 Perseus  (Per)\color{black}, Rd14 & $24$ &   $0.018$ & $0.0\color{black}61$ & $0.35\pm0.05$\\[0.2cm]
 Outer (Out)\color{black}, Rd14 & $6$ & $0.012^{\vphantom{{T^T}^T}}$  & $0.070$  &  \\[0.2cm]
 \hline
 All objects (All)\color{black}, {Rd14} & $103$ & $0.025$ & \color{black}$0.069$ & $-$ \\[0.2cm]
 \color{black}	HMSFR masers, {Rs17} & \color{black}$112$ & \color{black}$0.025$ & \color{black}$0.073$ & $-$\\[0.2cm]
 \color{black}Non-HMSFR masers, \color{black}Rs17 & \color{black}$29$ &  \color{black}$0.037$ & \color{black}$0.063$ & $-$\\[0.2cm]
 \hline
\multicolumn{5}{l}{}\\ [-3mm]
\multicolumn{5}{p{15.3cm}}{
$N$ is the number of objects; $\me \sigma_{\varpi}$ and $\me (\sigma_{\varpi}/\varpi)$ are the median absolute and relative parallax errors; $\sigma_{\text{w}}$ is the natural rms scatter
of objects across the segment inferred in NV18 for the final estimate of $R_0 = 8.8~$kpc.}
 \end{tabular}
\end{table}

Note that the HMSFR masers added in Rs17
hardly change the median parallax uncertainties for
objects of this class (Table~1) and are located on
the  $(r,\sigma_\varpi)$ plane in good agreement with the Rd14
masers (Fig.~2a). $\me
\sigma_{\varpi}=0.037_{-0.007}^{+0.033}$  for the non-HMSFR masers is higher than $\me
\sigma_{\varpi}=0.025_{-0.003}^{+0.004}$
for the HMSFR masers from the Rs17 catalogue
by a factor of $1.5$. Although these medians formally differ insignificantly, the overwhelming majority
of non-HMSFR sources (25 of 29) have~$\sigma_\varpi$ equal
to or larger than the median one for the HMSFR
masers (Fig.~2a). The slightly smaller~$\me (\sigma_{\varpi}/\varpi)$  for
the non-HMSFR sources than that for the HMSFR
ones (Table~1) is unrepresentative, because the former
are predominantly near the Sun (Fig.~2b). In what
follows, we will rely on the results for the HMSFR
masers forming more homogeneous samples with
more accurate parallax measurements.

To check the significance of these trends, we estimated the linear correlation coefficients $\kappa$ between
the parallax uncertainties, on the one hand, and the
distances or parallaxes, on the other hand, for each
individual segment and for all HMSFR masers of
the Rd14 and Rs17 catalogues (Table~2). For each
estimate of the coefficient $\kappa(p_1,p_2)$ from the sample
values of $N$ pairs of random variables $p_1$ and $p_2$ we
calculated the pro\-ba\-bi\-lity~$P_{\kappa}$ to obtain a correlation
coefficient larger in absolute value than $|\kappa(p_1,p_2)|$
under the assumption that $p_1$ and $p_2$ were uncorrelated using the formula for small $N$
\begin{equation}\label{P_corr}
 P_{\kappa} = 1 - S_{N-2}(t), \quad t = \kappa(p_1,p_2)\sqrt{\frac{N-2}{1-\kappa^2(p_1,p_2)}}\,,
\end{equation}
where  $S_{N-2}(t)$ is Student's distribution with $N-2$ degrees of freedom (Press et al. 1997). $P_{\kappa}$ are
also given in Table~2. The null hypothesis about the
absence of a correlation may be rejected at $P_{\kappa} < 0.05$.
The values of $\kappa$ that are significant in accordance with
this rule are highlighted in Table~2 in boldface.

The results presented in Table~2 show that the correlations between $\sigma_{\varpi}$ and $r$ (negative) and between
$\sigma_{\varpi}$ and  ${\varpi}$ (positive) are significant not only for the
complete samples of HMSFRs, but also within the
segments of the Local and Sagittarius arms separately. Since the tendency for $\sigma_{\varpi}$  to drop with $r$ is
most pronounced at distances $r\la 2.5$~kpc (Figs.~1a
and~2a), it is less evident in the case of other segments, which occupy little this region (Perseus and
Scutum arms) or do not pass through it at all (Outer
arm) (Fig.~1a). However, when the two most distant
masers of the Scutum arm segment are excluded,
$\kappa{(\sigma_{\varpi}, r)}$ and $\kappa{(\sigma_{\varpi},
\varpi)}$ become significant for this
feature as well. The closeness of the significant
correlation coefficients to one another ($\kappa{(\sigma_{\varpi}, r)}=-0.55\div-0.4\color{black}4$,
$\kappa(\sigma_{\varpi},
\varpi)=+0.{\color{black}38}\div+0.60$) suggests that in all these cases we are dealing with the
same effects. In the opinion of Reid (2014), these may
include: (1) the decrease in the apparent size of maser
spots with distance, as a consequence of which the
accuracy of astrometric measurements, other things
being equal, increases; and (2) using a larger number
of observations for some distant objects (e.g., W~49).
In any case, the results obtained suggest that at the
present epoch the accuracy of maser parallax measurements, on average, increases with distance and,
consequently, the $\sigma_{\varpi}=\text{const}$ model for a segment
is inconsistent with the present-day data. Note that
the same tendency is independently traced by non-HMSFR masers, only with a shift toward larger~$\sigma_\varpi$
(Fig.~2a).

\begin{table}[t!]
   \centering
   \small
   \tabcolsep0.08cm
   \caption{\label{corr}\rm Linear correlation coefficients of the absolute and relative parallax uncertainties with the distance and parallax
estimates for the HMSFR masers.}

   \vspace{5 pt}
   \tabcolsep0.07cm
   \begin{tabular}{l|cc|cc|cc|cc}
    \hline
   Arm/sample,
    & $\kappa{(\sigma_{\varpi}, r)^{\vphantom{T^{T^T}}}}$ & $P_{\kappa}$ & $\kappa{(\sigma_{\varpi}/\varpi, r)}$ & $ P_{\kappa}$
    & $\kappa(\sigma_{\varpi}, \varpi)$  & $P_{\kappa}$& $\kappa{(\sigma_{\varpi}/\varpi, \varpi)}$ & \hskip-0.1cm $P_{\kappa}$\\
    \color{black}catalogue &&&&&&&&
    \\[0.15cm]
    \hline
    Sct\color{black}, Рд14 & $ -0.10$ & $0.69$ & $\bf +0.79$ & $ 0.0003^{\vphantom{T^{T^T}}}$           & $+0.23$ & $0.38$           & $\bf-0.65$   & \hskip-0.1cm$0.0055$  \\[0.25cm]
    Sct$^\text{a}$\color{black}, Rd14&${\bf -0.55}$ & $0.035          $ & $ +0.46$ & $ 0.076          $ & $\bf+0.52 $ & $0.048          $ & $-0.48 $ & \hskip-0.1cm$0.064          $ \\[0.3cm]
    Sgr\color{black}, Rd14 & ${\bf -0.55}$ & $0.018          $& $ +0.20$ & $0.43$              & $\bf+0.60$ & $0.0094$& $-0.26$   & \hskip-0.1cm$ 0.29 $            \\[0.3cm]
    Loc\color{black}, Rd14 & ${\bf-0.4\color{black}7}$ & $ 0.02\color{black}0$  & $\color{black} \bf{+0.55}$ & \color{black} $0.0047$&\color{black} $\bf+0.38$ &\color{black} $ 0.059          $ &\color{black} $\bf-0.48$ & \hskip-0.1cm\color{black}$0.016$             \\[0.3cm]
    Per\color{black}, Rd14 & $-0.3\color{black}3$ & $0.1\color{black}1$ & $+0.0\color{black}8$   &\color{black} $0.72$                       & $+0.3\color{black}5$ & $ 0.1\color{black}0$            &\color{black} $-0.18$ & \hskip-0.1cm\color{black}$ 0.41$               \\[0.3cm]
    Out\color{black}, Rd14 & $+0.06 $ & $ 0.90$ & $ +0.19$& $0.72$                       & $-0.15$ & $0.78$             & $-0.28^{\vphantom{I}}$&\hskip-0.1cm$0.60$ \\[0.3cm]
    \hline
    All\color{black}, {{Rd14}} & $\bf{-0.4\color{black}7}$ &\color{black} \!$8\cdot10^{-5}$ &$\bf{+0.4\color{black}1}$ &\color{black} $3\cdot10^{-5}$  &\color{black} $\bf{+0.48}$ &\color{black} $2\cdot10^{-7}$ & $\bf{-0.3\color{black}6}$ & \hskip-0.1cm\color{black}$0.0002^{\vphantom{T^{T^T}}}$ \\  [0.3cm]
    \color{black}HMSFRs, Rs17 &\color{black} $\bf{-0.44}$ &\color{black} \!$0.0001$  &\color{black} $\bf{+0.40}$ &\color{black} $0.0001$   & \color{black}$\bf{+0.49}$ & \color{black}$0.0002$ &\color{black} $\bf{-0.35}$  & \hskip-0.1cm\color{black}$0.0003$ \\  [0.2cm]
  \hline
    \multicolumn{9}{l}{}\\ [-3mm]
    \multicolumn{9}{l}{$\!^{\text{a}}$ Without two masers of the Scutum arm at  $r\geqslant 8$~kpc.}\\[0.1cm]
    \multicolumn{9}{p{16cm}}{$P_{\kappa} \equiv  P (|\varkappa| > |\kappa(p_1, p_2)|)$ is the probability to obtain a correlation coefficient 
 $\varkappa$ larger in absolute value than the measured
$|\kappa(p_1, p_2)|$ in the absence of a correlation between the random variables
 $p_1$ and $p_2$\,. The significant correlation coefficients
 ($P_{\kappa}<0.05$) are
highlighted in boldface.}\\
   \end{tabular}
   \end{table}

Given the trend $\langle\sigma_\varpi\rangle \propto r^{-1}$, one might expect
that the relative uncertainty $\sigma_\varpi/\varpi$, on average, does
not change with $r$. A direct comparison of $\sigma_\varpi/\varpi$
with $r$ (Figs.~1b and~2b) shows that the scatter of
$\sigma_\varpi/\varpi$ for the HMSFRs actually remains approximately constant at $r\la 3.2$~kpc. Outside this region,
as $r$ increases, the values of $\sigma_\varpi/\varpi$  shift, on the whole,
to large values and several measurements with a low
relative accuracy appear. However, this tendency is
much less pronounced than the reduction in $\langle\sigma_\varpi\rangle$
with $r$. Although for complete samples of HMSFRs
the correlations between $\sigma_\varpi/\varpi$ and $r$ (positive) and
between $\sigma_\varpi/\varpi$ and $\varpi$ (negative) turn out to be significant,
 for separate segments they are either insignificant or, for the Local and Scutum arms, are
significant only formally (Table~2). For example, for
the Scutum arm the significance of the correlations is
based only on the two most distant ($r\geqslant 8$~kpc) objects that differ sharply from other masers of the same
segment by large~$\sigma_\varpi/\varpi$, which exceed $\me (\sigma_{\varpi}/\varpi)$
both for the objects of the Scutum arm and for the
entire Rd14 catalogue by several times (Fig.~1b, Table~1). When these two (of 17) objects are excluded,
the null hypothesis is not rejected for the Scutum
arm either (Table 2). For the Local arm the formal
significance appeared only after the replacement of
the data for one (!) object (L 1206) in Rs17. (No such
situation arises in the case of correlations for~$\sigma_\varpi$.) The
relative scatter of~$\me (\sigma_{\varpi}/\varpi)$ for different segments
is noticeably smaller than that of $\me (\sigma_{\varpi})$ (Table~1).
Thus, although the correlation between~$\sigma_\varpi/\varpi$  and $r$
apparently exists, it manifests itself weakly within the
arm segments considered separately. Therefore, the
$\sigma_{\varpi}/\varpi=\text{const}$ model for a separate segment may be
taken as consistent, to a first approximation, with the
present-day data on masers.

Since one of the probable causes of the reduction
in $\langle\sigma_\varpi\rangle$ with $r$ (using a larger number of observations
for distant masers) will not be necessarily retained in
future, in our numerical simulations we consider both
variants of allowance for the parallax uncertainty. Out
of them, as will be shown below, the $\sigma_{\varpi}/\varpi=\text{const}$ 
model describes a more favorable situation for estimating $R_0$ from the spiral-segment geometry, while
the $\sigma_{\varpi}=\text{const}$ model describes a less favorable situation. From the presented results and reasoning one
might expect the behavior of $\langle\sigma_\varpi\rangle$ with $r$, most likely,
to remain intermediate between these two cases in
any further parallax measurements.

\subsection{Modeling Procedure}
\vskip0.0cm
Consider three {\em{representative}\/} points $M_1$, $M_2$, and $M_3$ belonging to the segment of a model spiral
with parameters $R_0$, $i$, $\lambda_0$, $\lambda_1^\text{s}$, and $\Delta\lambda$ (see Subsection~2.1). Let us choose the positions of the points so
that the adjacent points are equidistant from one another in Galactocentric longitude~$\lambda$. The longitudes
of points $M_j$ are then defined by the expression $\lambda_j =
\lambda_1^\text{s} + \Delta\lambda\,(2j-1)/6$,  $j=1,2,3$. Thus, the distance
between the adjacent points is~$\Delta\lambda/3$, the distance
from the extreme point ($M_1$ or $M_3$) to the nearest (in
longitude) segment boundary is $\Delta\lambda/6$. Each of points
$M_j$ represents the position of one third of the segment
in longitude. The equidistant configuration of the
representative points corresponds to a uniform distribution of objects in~$\lambda$, i.e., approximately uniform
along the segment. This model describes satisfactorily the reality in many cases (see, e.g., the figures in
Popova and Loktin (2005) and Dambis et al. (2015)).
Therefore (and in order not to increase the number of
problem parameters), here we will restrict ourselves
only to this type of configurations.

To take into account the scatter of objects across
the arm segment and the parallax uncertainty in our
numerical experiments, the positions of the representative points $M_j$, $j=1,2,3$, of the model spiral
were varied as follows. Each point $M_j$ was shifted
randomly by distance $\rho_j$ in one or the other direction
along the straight line perpendicular to the model
spiral at point~$M_j$\,; in this way we found point $M_j'$
with parallax~$\varpi_j'$ (Fig.~3). Then, point~$M_j'$ was shifted
along the line of sight~$SM_j'$ also randomly, which
determined a pseudo-random point ($M_j''$ in Fig.~3)
with parallax~$\varpi_j''\,$.

Once the set of points $M_1''$, $M_2''$, and $M_3''$ had been
obtained, we searched for the parameters of the spirals
passing through them in one turn. In this case (i.e.,
within the three-point method), the values of $R_0$ for
such spirals are found from the following equation
(see NV18):
\begin{equation}\label{basic_ln}
(\Lambda_3 - \Lambda_2)\ln{R_1} + (\Lambda_1 - \Lambda_3)\ln{R_2} + (\Lambda_2 - \Lambda_1)\ln{R_3} = 0,
\end{equation}
where the Galactocentric distances $R_j$ of points $M_j''$
are expressed via the Cartesian heliocentric coordinates $X_j$ and $Y_j$ of these points from the formula
\begin{equation}\label{R_i_gal}
R_j = \sqrt{R_0^2 +X_j^2  + Y_j^2 - 2R_0X_j}, \quad j=1,2,3,
\end{equation}
while the nominal Galactocentric longitudes $\Lambda_j$ of
points $M_j''$ ($-\pi \leqslant \Lambda_j < \pi$, see NV18) are defined by
the formulas
\begin{equation} \label{X_i_La_i}
\sin{\Lambda_j} = \frac{Y_j}{R_j},\quad \cos{\Lambda_j} = \frac{R_0 - X_j}{R_j}, \quad j = 1, 2, 3.
\end{equation}

As shown in NV18, for the triplet of points taken from
the spiral, i.e., for $M_1$, $M_2$, and $M_3$, Eq. (2) can have
one or two additional roots. Therefore, for the triplet of
pseudo-random points $M_1''$, $M_2''$, and $M_3''$ we can also
obtain more than one (up to three) roots of Eq.~(2).
In such cases, among the values of $R_0$ found for a
given triple we chose $R_0$ closest to its initial value
for the model spiral and it was considered to be the
solution for the triplet. In addition, as our numerical
experiments showed, for points $M_1''$, $M_2''$, and $M_3''$
Eq.~(2) may not have any roots. This means that the
turn of at least one spiral cannot necessarily be drawn
through three arbitrary points if the straight line on
which the spiral pole can be located is specified. Such
triplets were excluded from the subsequent analysis.

The value of $R_0$ found for the triplet $M_1''$, $M_2''$, and
$M_3''$ also uniquely defines two other parameters ($i$ and
$\lambda_0$) of the spiral (Eqs.~(12) and~(13) in NV18). However, at this stage we do not consider the properties
of the $i$ and $\lambda_0$ estimates as derived characteristics
strongly correlating with $R_0$ (see NV18).

For the specified positions of the representative
points $M_j$ and the chosen values of other parameters,
we produced a set of $N_{\text{MC}}$ triplets of pseudo-random
points  $\left\lbrace
M_{1,m\,}'',M_{2,m\,}'',M_{3,m\,}''\right\rbrace_{m=1}^{N_{\text{MC}}}$, where $N_{\text{MC}}$ was
taken in such a way that the number $N_{\text{sol}}$  of solutions
for the problem of determining $R_0$, i.e., the number
of triplets with a nonempty set of roots of Eq.~(2).
was 10 000. We assumed that $N$ objects with parallax
estimates having an rms uncertainty $\sigma_\varpi$ or a relative
uncertainty $\sigma_\varpi/\varpi$ were observed in the spiral segment with a standard scatter of object across the arm~$\sigma_{\text{w}}$. Each point $M_j$ models the~mean position of one
third of the segment. Therefore, the standard error
in the position of $M_j$ in a direction across the arm
may be approximately taken to be~$\sigma_{\text{w}}\big/\!\sqrt{N/3}$, while
the standard error in the parallax of the shifted point
$M_j'$ (see Fig.~3) may be taken to be $\sigma_{\varpi}(M_j')\big/\!\sqrt{N/3}$,
where $\sigma_{\varpi}(M_j')$ is the parallax uncertainty at point $M_j'$\,.
Hence, for the generation of points $M_{j,m}''$\,, $j=1,2,3$, $m=1,2,\:\ldots,\:N_{\text{MC}}$\,, we varied $\rho_{j,m}$ (the shift of $M_{j,m}'$
relative to $M_{j,m}$) according to a normal law with
zero mean and standard deviation $\sigma_{\text{w}}\big/\!\sqrt{N/3}$
and $\varpi_{j,m}''$ according to a normal law with mean
 $\varpi_{j,m}'$ and standard deviation $\sigma_{\varpi}\big/\!\sqrt{N/3}$ in the case of a
constant absolute parallax error for the segment or $\varpi_{j,m}'(\sigma_{\varpi}/\varpi)\big/\!\sqrt{N/3}$  in the case of a constant relative
parallax error.

\begin{figure}
\centerline{%
\epsfxsize=7cm%
\epsffile{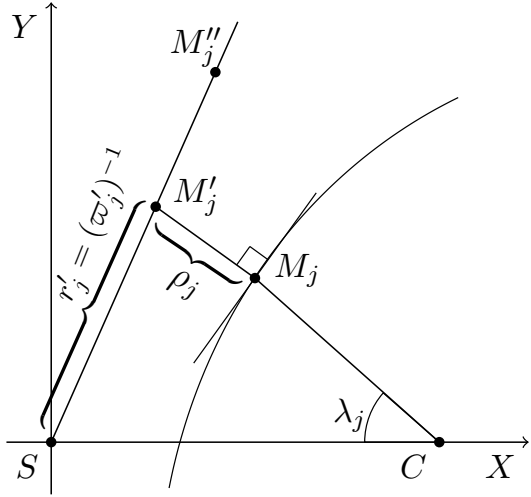}%
}%
\caption{\rm Varying the position of a representative point~$M_j$ along the perpendicular to the model spiral (point
 $M_j'$) and along the
line of sight (point~$M_j''$). $S$ is the Sun's position, $C$ is the pole of the model spiral (Galactic center), and 
 $(X,Y)$  is the Galactic plane.
} 
\label{represent_p}
\end{figure}

For the set of all our solutions $\lbrace R_{0,s}
\rbrace$, $s = 1,2,\:\ldots,\: N_{\text{sol\,}}$, we calculated the median $\me R_0$, the interval
$[\me R_0-\sigma^-_{R_0}, \me R_0+\sigma^+_{R_0}]$, with the probability for
the individual $R_{0,s}$~estimate to fall into the latter
being ${\approx}68.3\%$ ($1\sigma$ level), and the $1\sigma$ confidence
interval $[\me R_0-\sigma^-(\me R_0), \me
R_0+\sigma^-(\me R_0) ]$
for the median. We determined the quantities $\sigma^-_{R_0}$,  $\sigma^+_{R_0}$ and $ \sigma^-(\me R_0) $,
$\sigma^+(\me R_0)$ based on order
statistics for the ordered set of $R_{0,s}$~estimates (see,
e.g., Kobzar' 2006). We also calculated the mean
bias of the~$R_{0,s}$~estimates at given problem parameters: $\Delta R_0 = \me R_0 - R_0$ with the confidence interval $[\Delta R_0-\sigma^-(\me R_0),
\Delta R_0+\sigma^-(\me R_0) ]$, where
$R_0$ is the parameter of the original model spiral.

\subsection{Choosing the Basic Sets of Model Segment
Parameters}

Our numerical simulations were performed for
two families of model segments that represented the
Perseus and Scutum arms, respectively. These cases
approximately encompass the spread in probable
configurations of segments revealed at the present
epoch by objects with accurate heliocentric distances:
in the local Galactocentric sector (containing the
Sun) the Perseus arm passes outside the solar circle,
i.e., far from the Galactic center, while the Scutum
arm pases inside the circle $R = R_0$ relatively close
to the center (see Fig.~13 in NV18 and Fig.~4 in
this paper). In addition, the $R_0$ determination by the
three-point method in NV18 turned out to be reliable
precisely from masers in the Perseus and Scutum
arms. Therefore, it now seems most interesting to
investigate, also in prospect, the properties of the $R_0$
estimate from the geometry of these arm segments.

The parameter $R_0$ for the model segments was
taken to be $8$~kpc, in agreement with the recent mean
(''best'') values of ~$\langle
R_0\rangle_{\text{best}}=(7.9\div 8.3)
\pm (0.1\div 0.4)$~kpc (Nikiforov and Smirnova 2013; Bland-Hawthorn and Gerhard 2016; de Grijs and Bono 2016;
Camarillo et al. 2018) and the recent individual $R_0$
estimates (see, e.g., Rastorguev et al. 2017; Chen
et al. 2018; Majaess et al. 2018). The initial $R_0$
in this problem is not an important parameter and
did not change in our numerical experiments. For
other parameters of the model segments of each
of the two families we chose a set of such values
that exactly or approximately corresponded to the
characteristics found from the data of the catalogue
by Reid~et~al. (2014) for masers in the Perseus and
Scutum arms. We will call these sets and the values
of the parameters themselves in them {\em{basic}}. These
values are given in Table~3; the basic segments that
correspond to them are shown in Fig.~4. Other sets
of parameters for each family of models were formed
by varying one of the parameters, while the remaining
parameters retained their basic values.

The basic longitudes of the segment boundaries $\lambda_1^\text{s}$
and $\lambda_2^\text{s}$ correspond to the boundaries of the region occupied by the maser sources of a given arm. The basic
values of $i_0$ and $\lambda_0$ were chosen so that the model
segment passed through the region represented by
the arm masers.

For the Perseus arm the basic scatter across the
segment $\sigma_{\text{w}}$ was taken to be $0.35$~kpc, in accordance with the estimate that we 
obtained by applying the three-point method of determining the model
spiral parameters to the segment masers (NV18).
For the Scutum arm $\sigma_{\text{w}}=0.17$~kpc found by Reid~et~al. (2014) was taken for the same parameter.

The basic values of~$\sigma_{\varpi}$ and~$\sigma_{\varpi}/\varpi$ were chosen
to be equal to the medians $\me
\sigma_{\varpi}$ and $\me (\sigma_{\varpi}/\varpi)$ calculated for the masers of the corresponding arm
(Table~1).

The basic number $N$ of objects representing the
segment for the Perseus arm was taken to be equal to
the number of masers in the catalogue by Reid~et~al.
(2014) attributed to this arm. For the Scutum arm
this number, for convenience (because each representative point $M_j$ represents $N/3$ segment objects),
was taken to be 18, the number that is closest to
the number of arm masers ($N = 17$) in the same
catalogue and, at the same time, a multiple of three
(Table~1).

\begin{figure}[t!]
  \vspace{-1.2em}
\centerline{%
\epsfxsize=9.0cm%
\epsffile{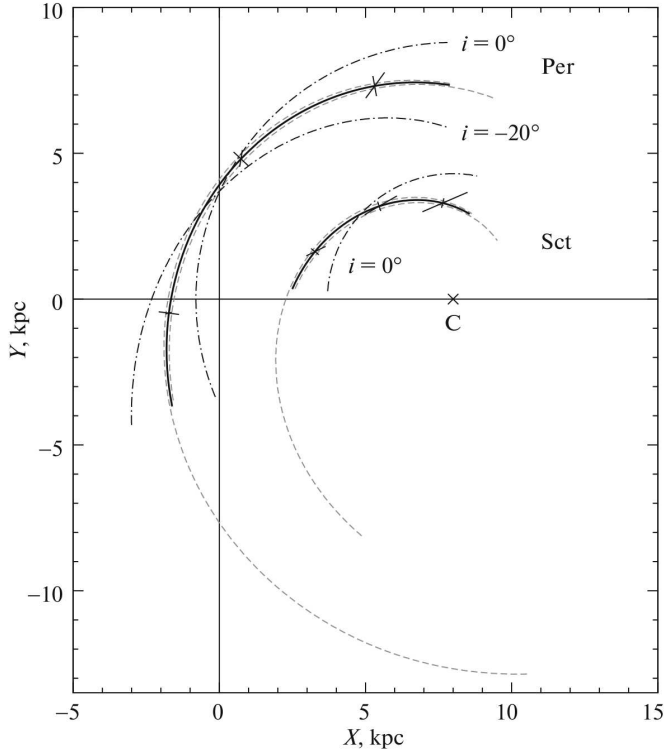}%
}%

\caption{\rm Scheme of the model spiral segments considered in our numerical experiments: the basic segments (black curves),
the segments of the largest angular extent
 $\Delta\lambda$ when fixing the boundary  $\lambda_1^\text{s}$ or
$\lambda_2^\text{s}$ (gray dashed curves, given with a slight
shift in radius), and the segments with the largest and smallest pitch angles
 $i$ (black dash--dotted curves). The bars indicate
the  $1\sigma$ distance uncertainty
 (at $\sigma_\varpi/\varpi=\text{const}$) and the scatter across the arm at three representative points~$M_j$ of the basic
segments. For point~$M_1$ of the Perseus arm the first of these bars exceeds only slightly the thickness of the model segment line.
Point $C$ is the pole of all segments (Galactic center), the Sun is at the coordinate origin.}
\label{model_segments}
\end{figure}

Note that the basic set of parameters for the
Perseus arm defines the segment crossing the Galactic center--anticenter line (Table~3, Fig.~4). In this
case, in most of the generated triplets of representative points $M_{1,m}''\,$, $M_{2,m}''\,$, and $M_{3,m}''$ the individual
points are located on different sides of this line. For
such a configuration, in the overwhelming majority
of cases, only one spiral passes through the triplet
of points (see NV18); therefore, there is no need to
choose the solution for $R_0$ from the set of roots of
Eq.~(2). The basic set for the Scutum arm specifies
the segment that is completely on one side of the
center--anticenter line, in quadrant~I (Table~3, Fig.~4).
In most cases, the generated triplets also turn out
to be there. Two spirals always pass through the
triplets arranged in this way for a nonempty set of
roots of Eq.~(2) and in nondegenerate cases (NV18).
Therefore, choosing the root closest to the initial $R_0$
of the model spiral from the roots of~(2) is necessary.
The cases where there are no roots for some generated
triplets (no solution for $R_0$) take place only for the sets
of parameters at which the dispersion of points $M_{1,m}''\,$, $M_{2,m}''\,$, and $M_{3,m}''$  relative to the model segment is
significant compared to the linear size and curvature
radius of the segment. For example, for the family of
Perseus arm models this occurs, very rarely, only at
$\Delta \lambda= 50\deg$, a value that is less than half the basic one.
For the family of the Scutum arm, on the contrary,
such cases are commonplace, because the model
segments of the family have considerably smaller
linear extent and curvature radius than those of the
Perseus arm segments (Fig.~4).

\begin{table}[t!]
 \centering
 \small
 \tabcolsep0.6cm
 \caption{\label{Per_Sct_par}\rm\!Basic sets of parameters for the model spiral segments representing the Perseus and Scutum arms in our numerical
experiments}
 \vspace{5 pt}
 \begin{tabular}{lrr||lll}
 \hline
 Parameter  & Per & Sct & Parameter  & Per & Sct \\
  \hline
   $R_0$, kpc  & $8.0$  & $8.0$             & $\sigma_{\varpi}$,~mas & $0.018$ & $0.028$ \\
   $i$  & $-10\deg$ &  $-20\deg$        &  $\sigma_{\varpi}/\varpi$ & $0.06$  & $0.10$ \\
   $\lambda_0$ & $+61\deg$  & $-55\deg$  & $\sigma_{\text{w}}$, kpc  & $0.35$ &  $0.17$\\
  \hline
  $\lambda_1^\text{s}$ & $-21^\circ$  & $+3^\circ$   & & & \\
  $\lambda_2^\text{s}$ & $+88^\circ$ & $+100^\circ$   &  $N$  & $24$ & $18$  \\
  $\Delta\lambda$ & $109^\circ$  & $97^\circ$         & & &   \\
  \hline
 \end{tabular}
\end{table}

\section{RESULTS}

The results of our numerical simulations, the  $1\sigma$
statistical uncertainty ($\sigma^+_{R_0}$, $\sigma^-_{R_0}$) and the mean bias
($\Delta R_0$) of the $R_0$ estimate from the spiral-segment geometry, for the basic sets of parameters (Table~3) and
when replacing the basic value of one of the dispersions with zero in these sets are presented in Table~4.
For the Perseus arm no significant biases~$\Delta R_0$ were
detected in all these variants of experiments. With
the basic sets for the Scutum arm~$\Delta R_0$ differ significantly from zero, substantially for the $\sigma_\varpi=\text{const}$
model. Comparison with the results for the same
arm when one of the dispersions is set equal to zero
(Table 4) shows that the bias is attributable to the
parallax uncertainty (at $\sigma_\varpi=0$~mas, $\Delta R_0=0$~kpc
within the error limits), but, on the other hand, the
scatter across the arm, in addition to the nonzero
$\sigma_\varpi$, increases $\Delta R_0$\,. The uncertainty in $R_0$ and the
contribution of the parallax uncertainty to it depend
strongly on the model for the latter within the arm:
at $\sigma_\varpi=\text{const}$ the values of $\sigma^\pm_{R_0}$ are larger than those
at $\sigma_\varpi/\varpi=\text{const}$ by 40--70\%. In the former case,
the contribution of~$\sigma_\varpi$ to the total uncertainty in $R_0$
clearly dominates (cf. the rows for $\sigma_\varpi=\text{const}\neq 0$ and $\sigma_\varpi=0$  in Table~4); in the latter case, the
contributions of both dispersions are, on the whole,
comparable, while for the Perseus arm the scatter
across the arm is even a more important factor (cf. the
rows for $\sigma_\varpi/\varpi=\text{const}\neq 0$ and $\sigma_\varpi=0$ in Table~4).
For the two arms considered the uncertainties in $R_0$
with the basic sets of parameters turned out to be
close in the case of $\sigma_\varpi/\varpi=\text{const}$; the uncertainty
in $R_0$ for the Scutum arm is slightly higher in the case
of $\sigma_\varpi=\text{const}$.

  \begin{table}[t!]
  \centering
  \small
 \caption{\label{basic}\rm Statistical uncertainty ($\sigma^+_{R_0}$, $\sigma^-_{R_0}$) and mean bias
 ($\Delta R_0$) of the~$R_0$ estimate from the spiral-segment geometry
for the basic set of the parameters for the Perseus and Scutum arm models and with one of the dispersions set equal
to zero.}
  \vspace{5 pt}
 \begin{tabular}{l|l|c|l||l|l|c|l}
 \hline
 \multicolumn{4}{c||}{Perseus arm} & \multicolumn{4}{|c}{Scutum arm}\\
 \hline
 $\sigma_{\varpi}/\varpi$   & $\sigma_{\text{w}}$\,,  &  $\sigma^\pm_{R_0}$\,,       & $\Delta R_0\,$, kpc  & $\sigma_{\varpi}/\varpi$   & $\sigma_{\text{w}}$\,,  &   $\sigma^\pm_{R_0}$\,,       & $\Delta R_0\,$, kpc  \\
 or $\sigma_{\varpi}$, mas & kpc                     &  kpc                     &                      & or $\sigma_{\varpi}$, mas & kpc& kpc &  \\
 \hline
 \rule{0pt}{1.2em}
 $\sigma_{\varpi}/\varpi = 0.06$ & $0.35$ & $^{+1.06}_{-0.95}$ & $-0.00{\color{black}6}_{-0.014}^{+0.015}$ & $\sigma_{\varpi}/\varpi = 0.1$ & $ 0.17$ & $^{+0.99}_{-1.03}$ &  $+0.116_{-0.014}^{+0.012}$ \\[0.1cm]
                                 & $ 0$   & $^{+0.54}_{-0.52}$ & $\phantom{+}0.000_{-0.006}^{+0.007}$ &                                & $0$ & $^{+0.84}_{-0.90}$ & $+0.044_{-0.012}^{+0.011}$ \\[0.1cm]
 \hline
 \rule{0pt}{1.2em}
 $\sigma_{\varpi} = 0.018$       & $0.35$ & $^{+1.55}_{-1.39}$ & $+0.00{\color{black}5}_{-0.019}^{+0.019}$  & $\sigma_{\varpi} = 0.028$&$ 0.17$ & $^{+1.71}_{-1.49} $ & $+0.562_{-0.019}^{+0.021}$ \\[0.1cm]
                                 & $0$    & $_{-1.16}^{+1.23}$ & $+0.014_{-0.015}^{+0.018}$            &                          &$ 0$ & $^{+1.65}_{-1.45}$ & $+0.505_{-0.017}^{+0.023}$  \\[0.1cm]
 \hline
 \rule{0pt}{1.2em}
 $\sigma_{\varpi}= 0$            & $ 0.35$& $_{-0.80}^{+0.89}$ & $-0.012_{-0.010}^{+0.012}$ & $\sigma_{\varpi}= 0$ &$ 0.17$ & $^{+0.60}_{-0.58}$ & $-0.005_{-0.007}^{+0.008}$ \\[0.1cm]
   \hline
 \end{tabular}
  \end{table}

To investigate the dependence of the results on
significant problem parameters, we performed series
of numerical experiments in each of which one of the
parameters (except $R_0$ and $\lambda_0$) changed in a wide
neighborhood of its basic value, while other parameters retained their basic values. We will call these
series {\em{basic}}. When varying the angular extent $\Delta\lambda$ of
the segment, we fixed either its left ($\lambda_1^\text{s}$) or right edge
($\lambda_2^\text{s}$). In all cases, except the series for different $\sigma_\varpi$\,,
we adopted the distance uncertainty model $\sigma_\varpi/\varpi=\text{const}$, because this variant agrees better with the
present-day data on masers (see Subsection~2.2).
When varying the pitch angle $i$, $\lambda_0$ was chosen for
each $i$ to be such that the model spiral passed through
the region occupied by the masers of this arm from the
data by Reid~et~al. (2014). The model segments for
the limiting $i$ and the largest $\Delta\lambda$ are shown in Fig.~4.
To clarify the role of each of the two dispersions in
the problem, for some basic series we performed {\em{additional}\/} series of experiments under the assumption
of $\sigma_\varpi=0$ or $\sigma_{\text{w}}=0$. As a rule, this was done when
in the basic series there were cases of biases $\Delta R_0$\,
differing significantly from zero. The results obtained
are summarized in Tables~5 and~6 and are graphically
presented in Figs.~5--11.

In Tables~5 and~6 the characteristics of the distribution of $R_{0,s}$\, estimates are the same as those in
Table~4 and are given for the extreme values of the
parameters being varied for the Perseus and Scutum arm models, respectively. Figures 5--11 show
the dependences of the median~$\me R_0$ and interval $[\me R_0-\sigma^-_{R_0}, \me R_0+\sigma^+_{R_0}]$ (gray bars) on various
parameters for the~$R_{0,s}$ estimates obtained in the
basic and additional series for each of the two arms. In
this figures the thickened bar marks the result at the
basic value of the parameter being varied in a given
series, while the horizontal dashed line indicates the
initial~$R_0=8$~kpc.

The extent of the segment $\Delta \lambda$ turned out to be the
most important parameter of the problem (Tables~5,~6
and Figs.~5,~6). The uncertainty in $R_0$ increases with
decreasing $\Delta \lambda$, very sharply for the Perseus arm at
$\Delta \lambda\la 70\deg$~(Fig.~5). Such short segments of an outer
arm essentially impose no constraints on $R_0$\,. As
the Scutum arm segment is reduced, $\sigma^\pm_{R_0}$ increase
not so dramatically, but a significant bias~$\Delta R_0$ appears at $\Delta \lambda\la 50\deg$, which is attributable mainly to
the nonzero spiral arm thickness, i.e., which cannot
be removed by reducing the parallax errors (cf. the
corresponding rows in Table~6 and the panels in Fig.~5
at the basic values of the two dispersions and when
either one or the other dispersion is set equal to zero).
For a short segment close to the Galactic center the
scatter of objects relative to the center line is not
small with respect to the segment's linear extent even
at accurate distances and, therefore, its curvature is
established unreliably. However, at $\Delta \lambda \ga 60\deg$ segments close to the center can be used to estimate~$R_0$,
but only when the parallax uncertainty is necessarily
taken into account in the method to avoid possible
biases (the right panels in Fig.~5). At~$R_0$ larger than
the basic one the uncertainty in $R_0$ drops rapidly with
increasing~$\Delta\lambda$ for both arms, decreasing by a factor
of $3$ compared to the basic results (obtained at the
basic values of the parameters), when the segment's
extent reaches approximately half the spiral turn (Tables~5,~6 and Figs.~4,~6).

\begin{table}[t!]
 \centering
 \small
 \caption{\rm\label{Perseus_res}Statistical uncertainty 
 ($\sigma^+_{R_0}$, $\sigma^-_{R_0}$) and mean bias
 ($\Delta R_0$) of the~$R_0$  estimate from the spiral-segment geometry
at the extreme deviations of one of the problem parameters from its basic value for the Perseus arm models}
 \vspace{5 pt}
 \begin{tabular}{l||c|ll||c|ll}
 \hline
   $p$ & $p_{\mathrm{min}}$ &  $\sigma^\pm_{R_0}$\,, kpc       & $\Delta R_0\,$, kpc      &  $p_{\mathrm{max}}$ &  $\sigma^\pm_{R_0}\,$, kpc        & $\Delta R_0\,$, kpc     \\
   \hline 
   $\Delta \lambda$, $\lambda_1^\text{s} = -21^\circ$ & $50^\circ$  & $ {}_{-2.7}^{+5.9}  $ & ${{-0.071}_{-0.048}^{+0.045}}^{\vphantom{T^{T^T}}}$ &  $120^\circ$ & ${}_{-0.89}^{+0.94}$ & $-0.004 \pm 0.013$\\ [0.1cm]  
   $\Delta \lambda$,  $\lambda_2^\text{s} = +88^\circ$  & $109^\circ$  & $ {}_{-0.94}^{+1.05}  $ & $-0.006_{-0.014}^{+0.015}$ &  $190^\circ$ & ${}_{-0.33}^{+0.34}$ & $\phantom{-}0.000 \pm 0.004$\\ [0.1cm] 
   $\sigma_{\varpi}/\varpi$ & $0.00$ & $ {}_{-0.79}^{+0.89}$ &  $-0.012_{-0.010}^{+0.012}$  & $0.20$ & $ {}_{-1.8}^{+2.1}$ & $-0.012_{-0.024}^{+0.034}$  \\[0.1cm] 
   $\sigma_{\varpi}$, mas & $0.00$ & $ {}_{-0.79}^{+0.89}$ & $-0.012_{-0.010}^{+0.012}$ & $0.05$ & $ {}_{-3.1}^{+4.1}$ & $ {-0.020}_{-0.036}^{+0.045}$ \\ [0.1cm] 
   $\sigma_{\mathrm{w}}$, kpc & $0.00$ & $ {}_{-0.52}^{+0.54}$ & $\phantom{+}0.000_{-0.006}^{+0.007}$ & $0.60$ &  $ {}_{-1.4}^{+1.7}$ & $-0.026_{-0.021}^{+0.022}$  \\[0.1cm] 
   $N$    & $3$ & $ {}_{-2.6}^{+3.4}$ &  $  {-0.022}_{-0.040}^{+0.030}$ & $60$ &  $ {}_{-0.61}^{+0.66}$ &  $-0.004_{-0.009}^{+0.010}$\\ [0.1cm] 
   \hline
    $i$ & $-20^\circ$  & $ {}_{-0.85}^{+0.94}$& $+0.001_{-0.013}^{+0.008}$  & $0^\circ$ & $ {}_{-1.01}^{+1.14}$  & $-0.014_{-0.014}^{+0.017}$\\ [0.1cm] 
    $i$, $\sigma_{\varpi} = 0$ & $-20^\circ$  & $ {}_{-0.69}^{+0.77}$& $-0.008_{-0.010}^{+0.009}$  & $0^\circ$ & $ {}_{-0.89}^{+1.01}$  & $-0.021_{-0.013}^{+0.015}$\\ [0.1cm] 
    $i$, $\sigma_{\text{w}} = 0$ &  $-20^\circ$  & $ {}_{-0.49}^{+0.52}$& $-0.001_{-0.007}^{+0.006}$ &$0^\circ$ &$_{-0.48}^{+0.49}$ &  $+0.002_{-0.006}^{+0.007}$ \\[0.1cm]
   \hline
\multicolumn{7}{l}{}\\ [-3mm]
\multicolumn{7}{p{15.3cm}}{$p$~is the parameter being varied; $p_{\text{min}}$, $p_{\text{max}}$ are the minimum and maximum values of the parameter~$p$. In the presence of additional
conditions for our numerical experiments, they are listed in the first column.}
 \end{tabular}
\end{table}

    \begin{table}[h!]
 \centering
 \tabcolsep0.18cm
 \caption{\rm\label{Scutum_res}Same as Table 5 for the Scutum arm models.}
 \vspace{5 pt}
 \begin{tabular}{l||c|ll||c|ll}
 \hline
   $p$ & $p_{\mathrm{min}}$ &  $\sigma^\pm_{R_0}$\,,      & $\Delta R_0\,$,      &  $p_{\mathrm{max}}$ &  $\sigma^\pm_{R_0}\,$,        & $\Delta R_0\,$,     \\
 &   & kpc & kpc  & & kpc & kpc \\
   \hline 
   $\Delta \lambda$, $\lambda_1^\text{s} = +3^\circ$ & $50^\circ$  & $ {}_{-1.8}^{+2.3}  $ & $-0.47_{-0.03}^{+0.03^{\vphantom{T^{T}}}}$ &  $120^\circ$ & ${}_{-0.93}^{+0.92}$ & $+0.294_{-0.012}^{+0.014}$\\ [0.1cm] 
   $\Delta \lambda$, $\lambda_1^\text{s} = +3^\circ$, $\sigma_{\varpi} = 0$ &    $50^\circ$ & ${}_{-1.5}^{+2.1}$ & $-0.31_{-0.02}^{+0.02}$ & $120^\circ$ & ${}_{-0.49}^{+0.44}$ & $+0.001_{-0.005}^{+0.006}$   \\[0.1cm]
   $\Delta \lambda$, $\lambda_1^\text{s} = +3^\circ$, $\sigma_{\text{w}} = 0$ &  $50^\circ$ & ${}_{-1.2}^{+1.1}$ & $+0.051_{-0.019}^{+0.017}$ & $120^\circ$ & ${}_{-0.90}^{+0.87}$ & $+0.249_{-0.013}^{+0.010}$ \\[0.1cm]
    \hline
  $\Delta \lambda$,  $\lambda_2^\text{s} = +100^\circ$  & $97^\circ$  & $ {}_{-1.02}^{+0.99}  $ & $+0.11 \pm 0.01$ &  $170^\circ$ & ${}_{-0.29}^{+0.31}$ & $+0.008\pm 0.004$\\[0.1cm] 
   $\Delta \lambda$,  $\lambda_2^\text{s} = +100^\circ$, $\sigma_{\varpi} = 0$   & $97^\circ$  & $ {}_{-0.58}^{+0.60}$ & $-0.005_{-0.007}^{+0.008}$ & $170^\circ$ & ${}_{-0.17}^{+0.18}$ & $-0.002_{-0.002}^{+0.003}$   \\[0.1cm]
    $\Delta \lambda$,  $\lambda_2^\text{s} = +100^\circ$, $\sigma_{\text{w}} = 0$ & $97^\circ$  & $ {}_{-0.90}^{+0.84}$ & $+0.044\pm0.012$ & $170^\circ$ & ${}_{-0.23}^{+0.26}$ & $+0.003_{-0.003}^{+0.004}$\\[0.1cm]
    \hline
   $\sigma_{\varpi}/\varpi$ & $0.00$ & $ {}_{-0.58}^{+0.60}$ & $-0.005_{-0.007}^{+0.008}$ & $0.20$ & $ {}_{-1.4}^{+1.5}$ & $+0.48 \pm 0.02$  \\[0.1cm] 
      $\sigma_{\varpi}/\varpi$,  $\sigma_{\text{w}} = 0$ & $0.01$ & $_{-0.095}^{+0.092}$ & $\phantom{+}0.000\pm 0.001$ & $0.20$ & ${}_{-1.3}^{+1.5}$ & $+0.43_{-0.02}^{+0.02}$ \\[0.1cm]
    \hline
     $\sigma_{\varpi}$, mas & $0.00$ & $ {}_{-0.58}^{+0.60}$ & $-0.005_{-0.007}^{+0.008}$ & $0.05$ & $ {}_{-2.5}^{+2.5}$ & $ {+0.69}\pm0.03$ \\ [0.1cm] 
     $\sigma_{\varpi}$, mas, $\sigma_{\text{w}} = 0$ & $0.005$ & ${}_{-0.39}^{+0.36}$ & $-0.002_{-0.004}^{+0.005}$ & $0.05$ & ${}_{-2.4}^{+2.3}$ & $+0.67\pm0.03$ \\[0.1cm]
    \hline
    $\sigma_{\mathrm{w}}$, kpc & $0.00$ & $ {}_{-0.90}^{+0.84}$ & $+0.044_{-0.012}^{+0.011}$ & $0.60$ &  $ {}_{-1.6}^{+2.1}$ & $ +0.37 \pm 0.02 $  \\[0.1cm] 
    $\sigma_{\mathrm{w}}$, kpc, $\sigma_{\varpi} = 0$   & $0.05$ & ${}_{-0.17}^{+0.17}$ & $-0.002\pm0.002$ & $0.60$ & ${}_{-1.5}^{+2.0}$ & ${+0.28} \pm 0.02$ \\[0.1cm]
    \hline
  $N$    & $3$ & $ {}_{-1.8}^{+2.1}$ &  $  {+0.71}_{-0.02}^{+0.02}$ & $60$ &  $ {}_{-0.64}^{+0.58}$ &  $-0.002_{-0.008}^{+0.007}$\\ [0.1cm] 
  $N$,  $\sigma_{\varpi} = 0$ &   $3$ & ${}_{-1.2}^{+1.5}$ & $+0.17\pm0.02$ & $60$ & ${}_{-0.32}^{+0.33}$ & $-0.004\pm0.004$ \\[0.1cm]
     $N$, $\sigma_{\text{w}} = 0$ & $3$ & ${}_{-1.5}^{+1.7}$ & $+0.58\pm0.03$ & $60$ & ${}_{+0.49}^{-0.54}$ & $-0.007\pm0.006$\\[0.1cm]
    \hline
    $i$ & $-20^\circ$  & $ {}_{-1.03}^{+0.99}$  & $+0.11 \pm  0.01$  & $0^\circ$ &  $ {}_{-1.5}^{+1.5}$& $+0.33 \pm 0.02$\\ [0.1cm]
    $i$, $\sigma_{\varpi} = 0$ & $-20^\circ$  & $ {}_{-0.58}^{+0.60}$  & $-0.005 \pm  0.01$  & $0^\circ$ &  $ {}_{-0.88}^{+0.86}$& $+0.064 _{-0.015}^{+0.013}$\\ [0.1cm]
    $i$, $\sigma_{\text{w}} = 0$ & $-20^\circ$ & $_{-0.90}^{+0.84}$ & $+0.044_{-0.012}^{+0.012}$ & $0^\circ$ & $ {}_{-1.3}^{+1.4}$ & $+0.22\pm0.02 $ \\ [0.1cm]
    \hline
 \end{tabular}
 \end{table}

The second most important factor is the law of
change in the statistical measurement error of the
distance as one recedes from the Sun. Of course, as
the parallax uncertainty increases, in any case, the
dispersion of the $R_{0,s}$ estimates also grows, but the
rate of this growth depends strongly on the model
adopted for $\sigma_\varpi$ (Fig.~7, the panels for the basic series
in Fig.~8, and Tables~5,~6). Analysis of the results obtained showed that $\sigma^\pm_{R_0}$ in all of the cases considered
approximately follow a power law:
\begin{equation}\label{varsigma}
 \sigma_{R_0} (\varsigma/\varsigma_{\text{b}}) = k (\varsigma/\varsigma_{\text{b}})^\alpha + \sigma_{R_0} (\varsigma=0),
\end{equation}
where  $\varsigma$ is a parallax dispersion characteristic ($\sigma_\varpi$
or $\sigma_\varpi/\varpi$), $\varsigma_{\text{b}}$ is the basic value of~$\varsigma$; the index ,,$\pm$``
in Eq.~(5) is omitted for simplicity. The exponent~$\alpha$
turned out to be virtually independent of the model
for~$\sigma_\varpi$; it is apparently specified by the segment's
configuration: $\alpha\approx1.5$ for the Perseus arm and~$\alpha\approx 1$ for the Scutum arm. However, the coefficient~$k$
defining the~$\sigma_{R_0}$ growth scale for both arms in the
$\sigma_\varpi=\text{const}$ model is several times larger ($k\approx0.6$ for
the Perseus arm and 1 for the Scutum arm) than that
for $\sigma_\varpi/\varpi=\text{const}$ ($k\approx0.2$ and $0.4$, respectively). As
a result, for example, the ratio $\sigma_{R_0} (2)/\sigma_{R_0} (0)$ is ${\sim}1.5$
for the Perseus arm and ${\sim}2.5$ for the Scutum arm
at $\sigma_\varpi/\varpi=\text{const}$, while at $\sigma_\varpi=\text{const}$ it is ${\sim}3$ and
${\sim}4.5$, respectively. On the other hand, the complete
absence of errors in the parallaxes increases the statistical accuracy of the $R_{0,s}$ estimate compared to the
basic result at $\sigma_\varpi=\text{const}$ much more dramatically
(almost by a factor of $3$ for the Scutum arm) than it
does at $\sigma_\varpi/\varpi=\text{const}$ (see Table~4).

The dispersion of the $R_{0,s}$ estimates expectedly
decreases with increasing number $N$ of objects representing the segment; in the case of the Perseus arm,
and for $N/3 \geqslant 8$ in the case of the Scutum arm, the
agreement with the law
\begin{equation}\label{sigmaR0(N)}
\sigma_{R_0}^{\pm}(N)\propto N^{-1/2}
\end{equation}
is almost perfect (Fig.~9, upper panels). Poorer agreement for smaller $N$ in the case of the second arm is
obtained in the presence of a significant bias $\Delta R_0$\,.
The latter is caused mainly by the errors in the parallaxes (cf.\ the two lower and upper right panels in
Fig.~9). At $\sigma_\varpi=0$ mas close agreement with the
law~(6) is obtained already at  $N/3 \geqslant 3$ (the lower left
panel in Fig.~9). The drop in $\sigma^\pm_{R_0}(N)$ according to
the law~(6) and the absence of significant biases~$\Delta R_0$
at least at $N \geqslant 24$ (and for the Perseus arm also
at small~$N$) provide evidence for the consistency of
the $R_0$ estimate from the spiral-segment geometry
even in the case of applying a simplified (three-point)
method (when $N\to\infty$ the mean $\mex R_{0,s} \to R_0$ and
the dispersion $\var R_{0,s}\to 0$).

The values of $\sigma_{R_0}^{\pm}$ grow with increasing~$\sigma_{\mathrm{w}}$, along
with increasing~$\sigma_\varpi$ and $\sigma_\varpi/\varpi$, approximately according to a power law (Fig.~10). The role of the~$\sigma_{\mathrm{w}}$ contribution to the total uncertainty of the~$R_{0,s}$
estimate depends on the parallax accuracy level and
the form of the dependence~$\sigma_\varpi(r)$. For example, for
the Perseus arm, when the scatter~$\sigma_{\mathrm{w}}$ is set equal to
zero, $\sigma_{R_0}^{\pm}$ decrease by half for~$\sigma_\varpi/\varpi=\text{const}$ and only
by  ${\approx}\onequarter$  for $\sigma_\varpi=\text{const}$ (Table 4, the ''Per'' panel in
Fig.~10). For the Scutum arm the contribution of
the parallax errors dominates in both models for~$\sigma_\varpi$\,.
Therefore, setting~$\sigma_{\mathrm{w}}$ equal to zero in both cases
reduces~$\sigma_{R_0}^{\pm}$ insignificantly at the basic values of~$\sigma_\varpi$
and~$\sigma_\varpi/\varpi$ (Table~4, the~''Sct'' panels in Fig.~10).

  \begin{figure}[t]
  \vspace{-1.2em}
\centerline{%
\epsfxsize=14cm%
\epsffile{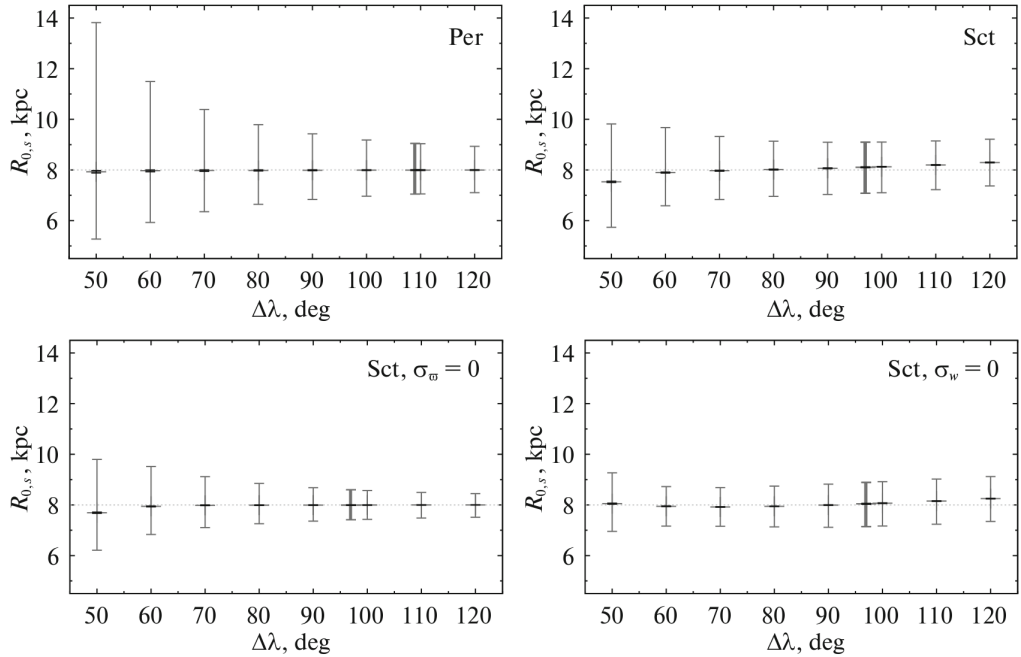}%
}
\vspace{-1em}

\caption{\rm\!Median and statistical uncertainty (the
 $1\sigma$ interval $[\me R_0-\sigma^-_{R_0}, \me R_0+\sigma^+_{R_0}]$, gray bars) of the~$R_0$ estimates from
the spiral-segment geometry as a function of the segment's angular extent in the case of a constant longitude of its left edge
 ($\lambda_1^\text{s}$) for the Perseus and Scutum arm models. On each panel the thickened gray bar marks the result at the basic value of the parameter being varied. The small black bars indicate the $1\sigma$ uncertainty of the median 
 $\me R_0$\,.}
\label{Per_Sct_dla}
\end{figure}

\begin{figure}[t!]
  \vspace{-1.2em}
\centerline{%
\epsfxsize=14cm%
\epsffile{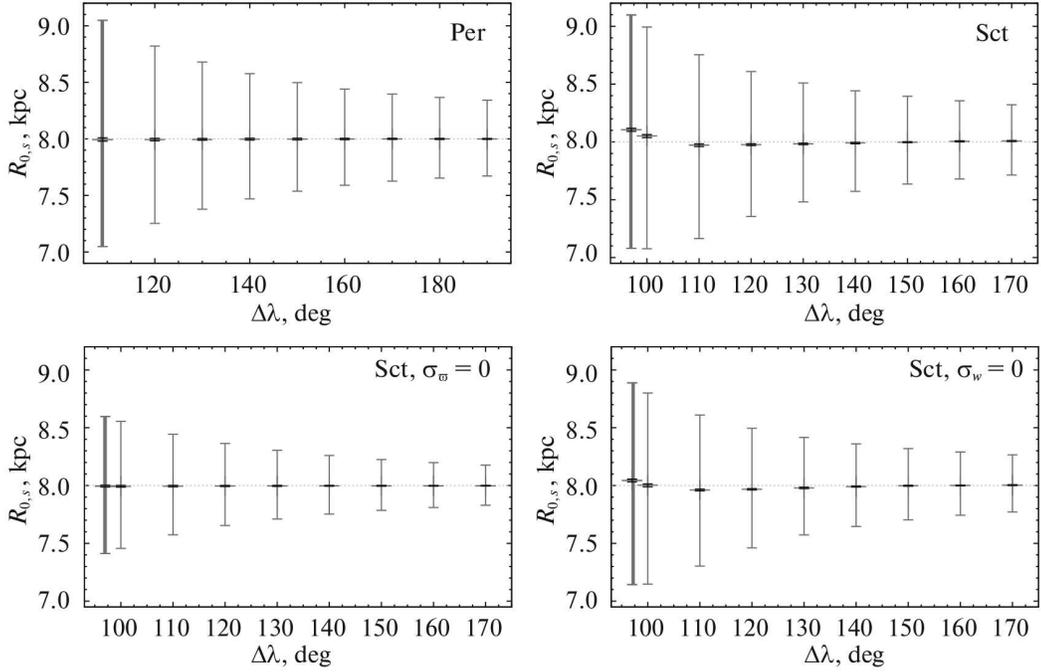}%
}
\vspace{-1em}

\caption{\rm\!Same as Fig.~5 in the case of a constant longitude of the segment's right edge~($\lambda_2^\text{s}$).}
\label{Per_Sct_dla1}
\end{figure}

\begin{figure}[t!]
  \vspace{-0.2em}
\centerline{%
\epsfxsize=14cm%
\epsffile{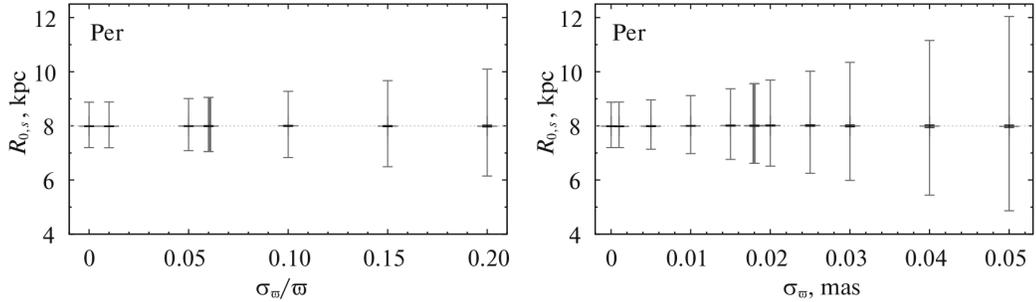}%
}
\vspace{-1em}

\caption{\rm\!Same as Fig.~5, but as a function of relative and absolute rms parallax errors for the Perseus arm models.}
\label{Per_sp}
\end{figure}

\begin{figure}[t!]
  \vspace{-0.2em}
\centerline{%
\epsfxsize=14cm%
\epsffile{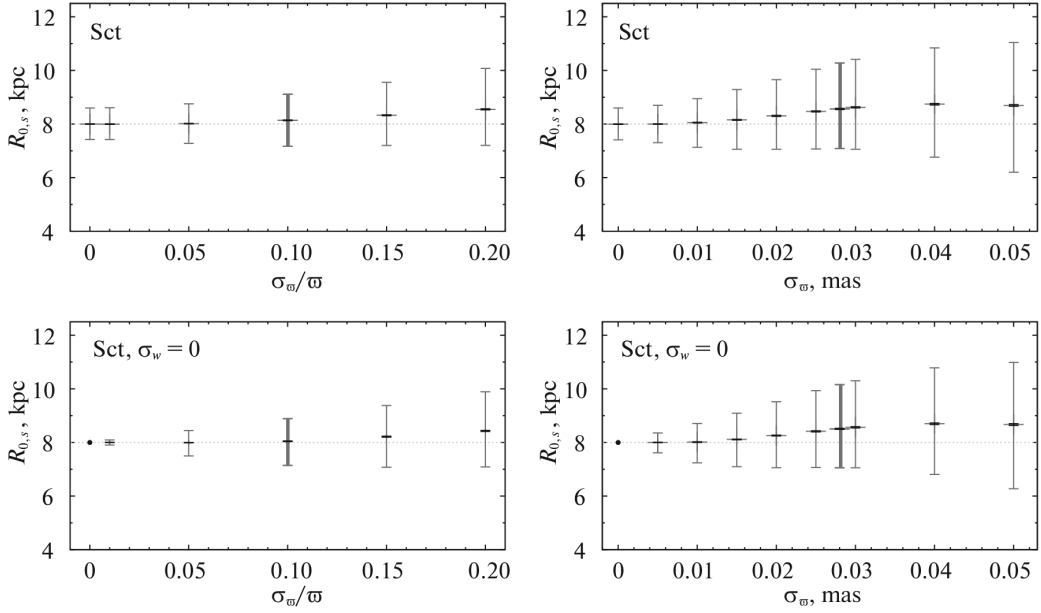}%
}
\vspace{-1em}

\caption{\rm\!Same as Fig.~7 for the Scutum arm models.}
\label{Per_Sct_spp}
\end{figure}

\begin{figure}[t!]
  \vspace{-0.2em}
\centerline{%
\epsfxsize=14cm%
\epsffile{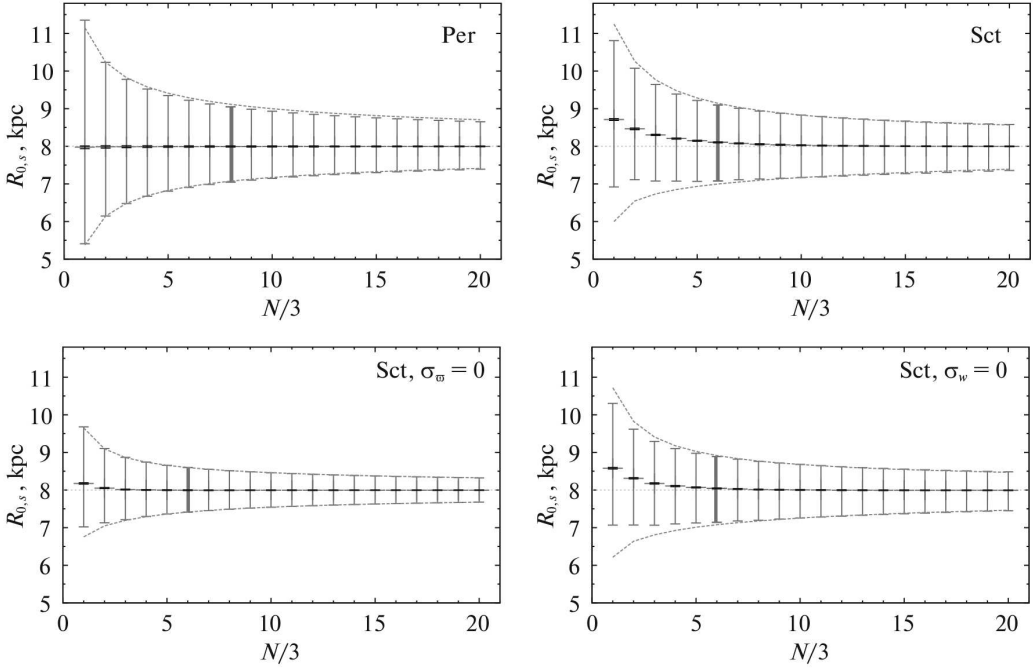}%
}
\vspace{-1em}
\caption{\rm\!Same as Fig.~5, but as a function of the number of objects populating the segment. \;\;\;\; The curves represent the laws
$\sigma_{R_0}^+(N)\propto 1/\sqrt{N}$ and $\sigma_{R_0}^-(N)\propto 1/\sqrt{N}$.}
\label{Per_Sct_N}
\end{figure}

\begin{figure}[t!]
  \vspace{-0.2em}
\centerline{%
\epsfxsize=14cm%
\epsffile{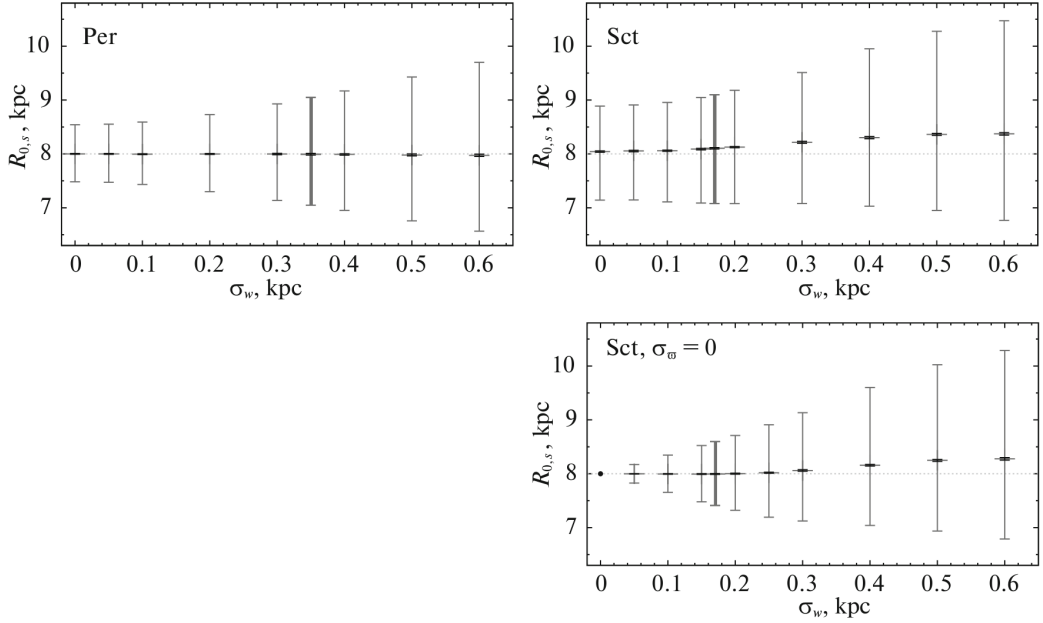}%
}
\vspace{-1em}
\caption{\rm\!Same as Fig. 5, but as a function of the scatter of objects across the model spiral segment.}
\label{Per_Sct_sw}
\end{figure}

\begin{figure}[t!]
  \vspace{-0.2em}
\centerline{%
\epsfxsize=14cm%
\epsffile{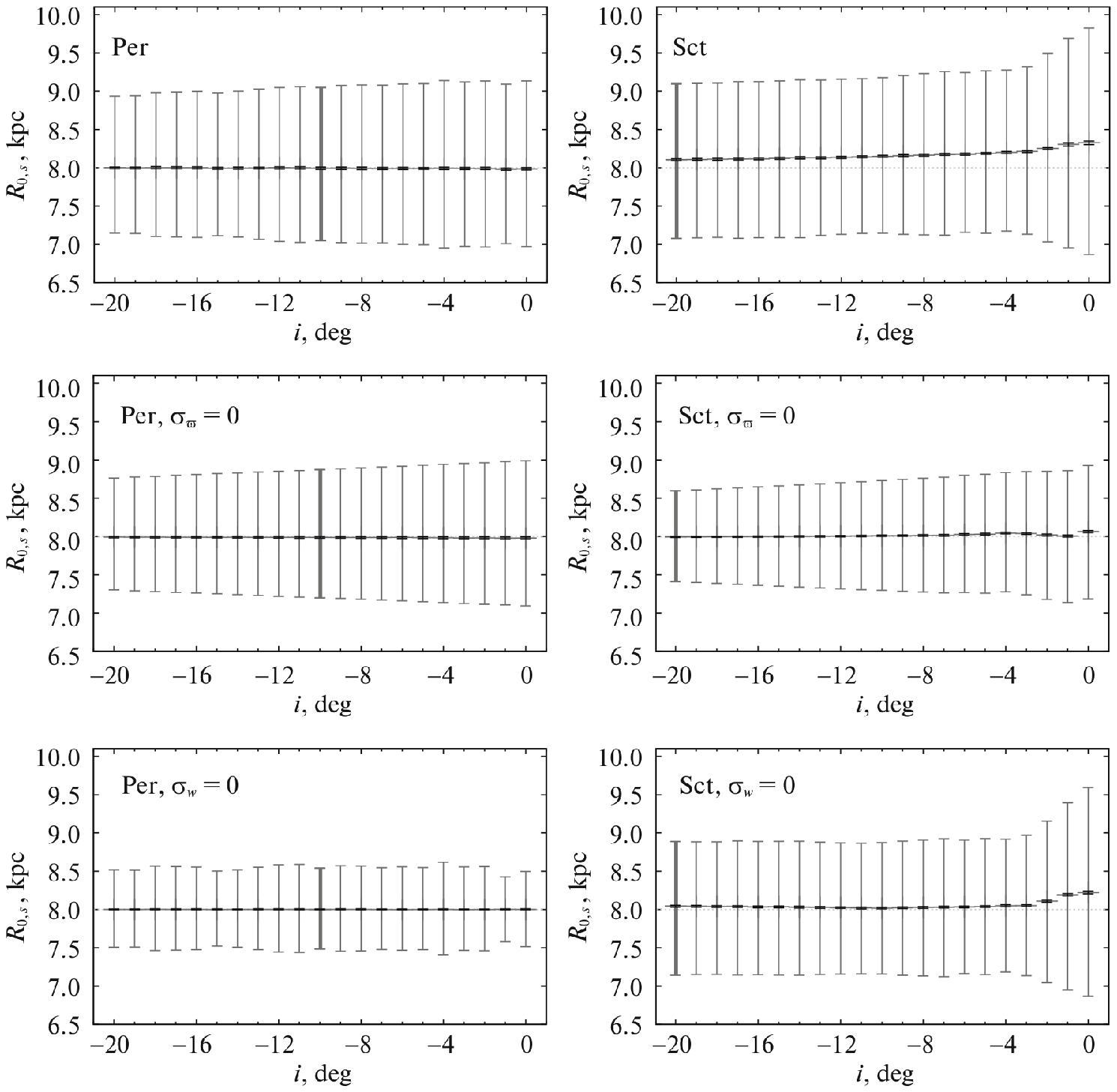}%
}
\vspace{-1em}
\caption{\rm\!Same as Fig. 5, but as a function of model spiral pitch angle.}
\label{Per_Sct_i}
\end{figure}

The dispersion of the~$R_{0,s}$ estimate is least affected
by the pitch angle $i$ of the model spiral (Fig.~11 and
Tables~5,~6). At the~basic values of both dispersions in
the problem the~general trend with changing $i$ is not
quite obvious (the upper panels in Fig.~11). However,
setting~$\sigma_\varpi$ equal to zero gives a clear picture: $\sigma^\pm_{R_0}$
grow almost linearly with increasing $|i|$ (the middle
panels in~Fig.~11). This means that in the absence
of measurement errors the distance from the Sun to
the center of the ring structure is determined from its
sector with a lower (!) accuracy than the distance
to the pole of the spiral arm from its segment at the
same angular extent of the sector and the segment.
The larger the arm pitch angle in absolute value, the
more accurate the result. The growth of $\sigma^\pm_{R_0}$ with
$|i|$ is moderate, but not negligible: at $\sigma_\varpi=0$ mas
in the interval $i\in[-20\deg,0\deg]$ it is a factor of ${\approx}1.3$ for
the Perseus arm and a factor of ${\approx}1.5$ for the Scutum
arm (Tables~5 and~6). For ideally thin structures
($\sigma_{\mathrm{w}}=0$ kpc) distorted by the errors in the parallaxes,
the behavior of~$\sigma^\pm_{R_0}(i)$ is determined by the scatter and
characteristic values of distances to the points of the
segment from the Sun and from the Galactic center,
the position of the segment relative to the center--anticenter axis, and other parameters. Therefore, for
different configurations the trends can be quite different (the lower panels in Fig.~11), whence the complex
dependences~$\sigma^\pm_{R_0}(i)$ under the combined action of
both dispersions (the~upper panels in Fig.~11).

\section{DISCUSSION}

Significant biases~$\Delta R_0$  were detected only for the
Scutum arm (Table~6, Figs.~5, 6, 8--11). This stems
from the fact that its segments have an appreciably
smaller linear extent and are much closer to the
Galactic center than the Perseus arm segments
(Fig.~4). As a~result, the solutions for the Scutum
arm turn out to be more sensitive to the dispersion
relative to the center line. In this case, the errors in
the parallaxes play a major role in the appearance of
significant biases: when they are set equal to zero,
these biases disappear in most cases (cf.\ the results
of the basic series and additional series at $\sigma_\varpi=0$
in the table and the figures mentioned above). The
biases~$\Delta R_0$ caused by the parallax uncertainty are
always positive (see the results at $\sigma_\text{w}=0$ there).
For a normal distribution of random parallax errors
the distribution of distance errors has a positive
skewness, leading to some overestimation of the distance, on average, which is greater for larger~$\sigma_\varpi$\,. In
combination with the location of the Scutum arm in
the general direction to the Galactic center, this leads
to an overestimation of~$R_{0,s}$ from the segments of this
arm, on average. However, the biases are not always
explained only by the errors in the parallaxes: at a
small extent ($\Delta \lambda\la 60\deg$), a sparse population ($N/3\leqslant 2$), and a large natural scatter ($\sigma_\text{w}\ga 0.3$~kpc) small,
but significant biases~$\Delta R_0=-0.3\div +0.3$ kpc can
result even for $\sigma_\varpi=0$~mas (Table~6; the ''Sct, $\sigma_\varpi=0$'' panels, Figs.~5, 9, 10). Their nature is different: if
the dispersion is not small compared to the segment
length, then for a significant fraction of the sets of
triplets~$M''_j$ (for example, lying on one straight line)
the information about the curvature of the initial
segment is completely lost; the formal solutions of~(2)
for such triplets lead to biases. To avoid noticeable
biases~$\Delta R_0$, we need: (1) to take into account the
errors in the parallaxes directly in the method and
(2) not to use short, sparsely populated, disperse
segments close to the Galactic center for our spatial
modeling.

Note that in our preliminary numerical experiments (Nikiforov and Veselova 2015) significant biases were also obtained for the Perseus arm in some
series. This was caused by the exclusion of the
triplets of points~$M''_j$ corresponding to leading spirals
($i>0\deg$) from consideration. Our new experiments
showed that, despite the seeming naturalness of this
constraint for our Galaxy, it actually cuts off part
of the distribution of~$R_{0,s}$ estimates and, thus, can
introduce a fictitious bias~$\Delta R_0$\,. When abandoning
the constraint~$i\leqslant 0\deg$, the biases for the Perseus arm
disappear. In this paper we did not impose this constraint.

In all series of experiments for the Perseus arm
and in most of the~series for the~Scutum arm it
turned out that $\sigma^+_{R_0}>\sigma^-_{R_0}$ (see the tables and figures
in Section~3). This is partly a consequence of the
skewness of the distribution of distance errors discussed above and partly a general geometric property
of some classes of methods for determining~$R_0$, i.e.,
the data agree better with the model at the trial $R'_0=\breve{R}_0+\delta R_0$\, than at $R'_0=\breve{R}_0-\delta R_0$\,, where $\breve{R}_0$ is the
optimal or true value, especially if $\delta R_0>0$~kpc is not
small compared to~$\breve{R}_0$ (see Nikiforov (1999, 2000) and
Nikiforov and Kazakevich (2009) as examples in the
kinematic class of methods).

The results obtained in the previous section show
that the dispersion of the~$R_0$ measurements from the
geometry of spiral segments depends significantly on
the law~$\sigma_\varpi(r)$. As an example, we generated pseudorandom ''observed'' distributions of objects (100 per
each degree of longitude~$\lambda$) belonging to one turn of a
spiral arm with parameters equal to the basic values
for the Perseus arm at a constant absolute parallax
uncertainty (Fig.~12a) and a constant relative parallax
uncertainty (Fig.~12b). These figures illustrate how
strongly the true distribution of spiral arm tracers for
the~$\sigma_{\varpi}=\text{const}$ model is smeared compared to the
~$\sigma_{\varpi}/\varpi=\text{const}$ model. The fact that the present-day data on masers are closer to the latter model
(Subsection~2.2) is a very lucky circumstance for the
prospects of applying this approach. In any case,
the errors in the parallaxes deform noticeably the
observed spatial distribution and, therefore, its proper
modeling is possible only when both dispersions are
taken into account. In some spiral arm segments a
particular dispersion dominates, while in other segments the dispersions are comparable in importance
(Figs.~12b and~12c). The results of our experiments
in Section~3 also indicate that, although, on the
whole, the distance measurement errors are a more
important factor, the contributions of the two dispersions to the total uncertainty in the $R_{0,s}$ estimate
are comparable (cf.\ the results at $\sigma_\varpi=0$~mas and
$\sigma_\text{w}=0$~kpc). This implies that none of the dispersions
can be ignored in our modeling.

\begin{figure}[t!]
  \vspace{-0.2em}
\centerline{%
\epsfxsize=10cm%
\epsffile{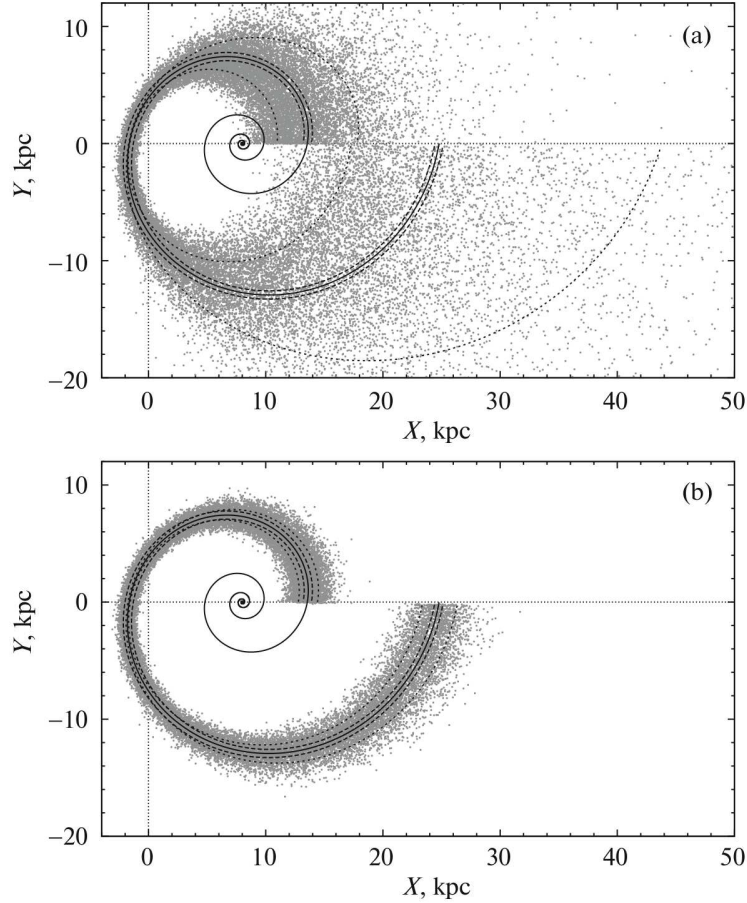}%
}
\vspace{-1em}
\caption{\rm\!Numerically generated distribution of 3600 objects belonging to one turn of a spiral arm with parameters of its center
line $R_0 = 8.0$~kpc, $i = -10^\circ$, $\lambda_0 = 61^\circ$
and a transverse scatter $\sigma_{\text{w}} = 0.35$~kpc at 
{\color{black}$\sigma_{\varpi} = 0.0175$~mas (a) and $\sigma_{\varpi}/{\varpi} = 0.06$ (b)}. The solid line indicates the model spiral, the dashed and dotted lines bound the region of deviation from the model spiral by $\pm 1\sigma_{\text{w}}$\,,  and by $\pm 1\sigma_{\varpi}$ (a) or $\pm 1\sigma_{\varpi}/\varpi$ (b), respectively.}
\label{spiral_sw_spp}
\end{figure}

For both arms the statistical uncertainty is $\sigma_{R_0}\approx 1$~kpc at the basic segment parameters (Table~4). As
has been shown in the previous section, the dispersion
of the~$R_{0,s}$ estimates is most strongly affected by
the extent~$\Delta \lambda$ of the segment. Therefore, it can be
assumed that the presence of edge objects, which, in
reality, specify the segment boundaries, must slightly
increase the conditionality of the~$R_0$ determination
problem compared to the results of modeling the segment by three representative points. Thus, one might
expect the~$\sigma_{R_0}$ estimates obtained in our experiments
to be upper limits. To estimate the overestimation
coefficient, we repeated the numerical experiments
in NV18 for masers distributed over the segment
(when abandoning the constraint on~$i$) and found
$\sigma^-_{R_0}=0.84$~kpc, $\sigma^+_{R_0}=0.85$~kpc for the Perseus arm
and $\sigma^-_{R_0}=0.73$~kpc, $\sigma^+_{R_0}=0.65$~kpc for the Scutum arm. Taking the largest of these quantities and
comparing it with the basic results in this paper, we
find that the overestimation of~$\sigma^\pm_{R_0}$ in our three-point
experiments is at least $15\%$ at parameters close to
the basic ones, i.e., as the uncertainty from {\em{only}
one\/} segment we may take a cautious estimate~$\sigma_{R_0}\simeq 0.85$~kpc. If the data on {\em{several}\/} ($N_\text{arm}$) segments
are used, then, other things being equal, one might
expect the final uncertainty
\begin{equation}
    \sigma_{R_0}\simeq\frac{0.85~\text{kpc}}{\sqrt N_\text{arm}}.
\end{equation}
For example, $\sigma_{R_0}\simeq
0.5$~kpc for three segments (open
clusters), $\sigma_{R_0}\simeq 0.4$~kpc for four segments (masers),
and $\sigma_{R_0}\simeq 0.3$~kpc for eight segments (classical
Cepheids). This is close to the statistical accuracy of
present-day~$R_0$ estimates (see, e.g., Bland-Hawthorn
and Gerhard 2016; de Grijs and Bono 2016).

Using short and sparsely populated segments as
well as intrinsically less accurate data on the heliocentric distances can reduce the predicted accuracy
of the solution. If the law $\sigma_\varpi=\text{const}$ is established
in future when measuring the parallaxes, then the
question about~$\sigma_\varpi$ will become acute. If, however,
the law~$\sigma_\varpi/\varpi=\text{const}$ is retained, then the relative
parallax accuracy will be not so important.

The accuracy of~$R_0$ can be improved most dramatically by increasing the extent of the identified
 segments. The next most important factors for refining
the solution are an increase in the number of objects
with independently measured distances attributed to
the segments and revealing new arms segments and
the internal structure of already known features in the
spiral pattern. The potential for reducing~$\sigma_{R_0}$ through
a more accurate parallax measurement (at~$\sigma_\varpi/\varpi=\text{const}$) is only moderate.

The presented results were obtained with a standard multiplicative congruent 
random number generator (IBM Manual C20-8011), which has an acceptable quality for our purposes in the two-dimensional
case. To check the stability of our results to the
choice of a generator, we also used one of the generators by Fishman and Moore (1986), the first in
their rating belonging to the same class for the most
dispersive series of experiments. With this generator
for the Perseus arm the biases~$\Delta R_0$ differ from those
presented in the paper within the error limits, while
the standards~$\sigma_{R_0}^\pm$ retain all of the presented signs.
For the Scutum arm with these two generators the
standards differ insignificantly within the entire series,
while the biases differ at moderate deviations of the
parameter being varied from the basic value. For the
extreme values of the parameter, at which the solution
for the Scutum arm is unstable for the above reasons,
with the second generator the biases turn out to be
slightly smaller, but remain significant in all cases
when they are significant with the first generator.
Thus, the application of a different random number
generator does not affect our conclusions.

On the whole, our results show that the geometric
estimation of~$R_0$ from spiral segments is operable for
a wide set of possible parameters even in the case
of a simplified (three-point method). However, the
capabilities of this, basically test method are limited:
it disregards the natural and measuring dispersions;
it is too sensitive to some peculiarities of the spatial
distribution of segment objects, especially at a small
sample size (see NV18); the solution can be directly
obtained with its help only for one segment and for
one type of model (logarithmic spiral). As was pointed
out in NV18, both dispersions can be taken into
account when implementing this approach within the
maximum likelihood method (MLM). Although this
method will be much more laborious, it will allow one
to estimate $R_0$ from the geometry of spiral segments
for arbitrary arm models and, simultaneously, to properly solve classical and new problems of modeling the
Galactic spiral structure.

Note that the possibility of introducing a more
general arm model is important for eliminating the
external error within this approach. In the case
of the three-point method, systematic deviations
from a logarithmic spiral can lead to such an error.
For example, if the representative points are taken
from an Archimedean spiral, then for the basic
segments of the Perseus and Scutum arms the
three-point method gives the biases $\Delta
R_0=-0.02$ 
and $+0.06$~kpc, respectively. However, according
to our preliminary estimates, significant deviations
from the shape of a logarithmic spiral cannot yet
be found from tracers with reliable photometric or
absolute distances for the arms of our Galaxy. Among
the various cases, the largest (in absolute value)
derivative of the pitch angle $i_1\equiv
\frac{di}{d\lambda}$ is $+0.029 \pm 0.076$
for the Perseus arm and $-0.04 \pm 0.25$ for the Scutum
arm\footnote{\color{black}Here $i$ and $\lambda$ are expressed in the same angular measure.} (both estimates differ insignificantly from zero).
If the representative points are taken from spirals with
a variable pitch angle $i(\lambda)=i_0+i_1(\lambda-\lambda_0)$, then
taking these values of $i_1$\,, this gives upper (in absolute
value) limits for the biases in the three-point method:
$\Delta R_0<+0.6$~kpc for the Perseus arm and $\Delta R_0>-0.4$~kpc for the Scutum arm. In all these cases,
the non-logarithmic spirals in the segment between
the representative points at the basic segment extent
differ little from the logarithmic models compared
to the uncertainty of the latter, while the biases for
individual segments are smaller or even appreciably
smaller than the statistical errors in~$R_0$ and have
opposite signs for different segments. Therefore, one
might expect the bias of the final~$R_0$ estimate from
several segments to be insignificant. In any case,
a bias of this type is not a fundamental problem
of our method for determining~$R_0$ provided that we
pass to the MLM that admits using more general
models. This will allow one to reveal systematic
deviations from a logarithmic spiral, if they exist,
and, simultaneously, to take them into account when
estimating the parameters. On the other hand,
the scale of potential biases for individual segments
shows that attention should be paid to searching for
such deviations.

\section{CONCLUSIONS}
To establish the capabilities and applicability
boundaries of the new method for determining the
distance to the Galactic center~$R_0$ proposed by us
previously (NV18), from the geometry of spiral arm
segments, we investigated the influence of various
factors on the statistical properties of the~$R_0$ estimate from a separate segment through numerical
simulations. The~$R_0$ estimates were found by a
simplified method reconstructing the geometry of a
segment from its three representative points, which
allowed a large number of numerical experiments
to be performed in a limited time. The problem
parameters were varied in wide neighborhoods of their
basic values characterizing the Perseus and Scutum
arm segments from the present-day data on masers
with trigonometric parallaxes.

The statistical uncertainty~$\sigma_{\varpi}$ in the present-day
parallax measurements for masers was shown to systematically decrease (!) with increasing heliocentric
distance~$r$, with the relative uncertainty~$\sigma_{\varpi}/\varpi$ remaining, on average, approximately constant (at least
at~$r\la 3.2$~kpc).

Our numerical experiments provide evidence for
the consistency of the~$R_0$ estimate from the spiral-arm geometry. Significant biases of the estimate
were detected only for the inner arm (Scutum); they
are attributable mainly to the random errors in the
parallaxes, which lead to an asymmetric distribution
of distances~$r$, as well as the small angular extent~$\Delta\lambda$
of the segment and the small number of $N$ of objects
representing it. The dispersion of the $R_0$ estimate
is affected most strongly by the extent $\Delta\lambda$ (as the
latter increases from the basic value to half the spiral
turn, $\sigma_{R_0}$ decreases by a factor of $3$). As the parallax
uncertainty grows, $\sigma_{R_0}$ increases. If in further parallax measurements~$\sigma_{\varpi}$, on average, remains constant
with~$r$, then~$\sigma_{\varpi}$ will be almost equally important for
the dispersion of~$R_0$\, as $\Delta\lambda$. When the law ~$\sigma_{\varpi}/\varpi=\text{const}$, which describes well the present-day data, is
retained, the remaining parameters, except the pitch
angle~$i$, exert an equally significant, but weaker influence on~$\sigma_{R_0}$. In the absence of parallax errors a
decrease in $|i|$ increases little the dispersion of $R_0$; in
particular, this implies that when the spiral segment
degenerates into a ring sector (at~$i = 0^\circ$), the distance
to the center of the latter is determined less accurately
than the distance to the pole of the spiral segment.

The applicability boundaries of the $R_0$ determination from the spiral-segment geometry are limited by
the dispersion of the~$R_0$ estimate ($\Delta\lambda>70\deg$, $\sigma_\varpi<0.04$~mas, $N/3>1$) for the outer arm (Perseus) and
by the presence of a significant bias in the $R_0$ estimate
caused not by the parallax errors, but by the finite
segment thickness ($\Delta\lambda>50\deg$, $\sigma_\text{w}<0.3$~kpc, $N/3>2$) for the inner arm (Scutum).

The fact that the present-day data on masers agree
better with the~$\sigma_{\varpi}/\varpi=\text{const}$ model is a very lucky
circumstance for applying our approach to these tracers of the spiral structure. The accuracy of the final~$R_0$ estimate can be improved by using several arm
segments in the analysis and by increasing the extent~$\Delta\lambda$ of the identified segments and the number
of objects with independent distance estimates attributed to them.

Our results suggest that the $R_0$ estimation from
spiral segments is operable for a wide set of possible
parameters even when using a robust, but inefficient
L-estimator (median) in the three-point method.
This makes the development of a more complex,
but more correct method based on an efficient M-estimator meaningful. As our numerical experiments
showed, the combined action of the measuring dispersion of distances and the natural scatter across
the arm generally leads to a more complex form of
the observed spatial distribution of objects, with the
influence of one of the dispersions on it, on the whole,
being comparable to the influence of the other one.
To properly model this distribution, the maximum-likelihood (ML) estimators can be used as an M-estimator.

\section*{ACKNOWLEDGMENTS}
We are grateful to the referees for their useful
remarks. This work was supported by the Russian
Science Foundation (grant no. 18-12-00050).

\section*{REFERENCES}

\begin{enumerate}

\item J. Bland-Hawthorn and O. Gerhard, Ann. Rev. Astron. Astrophys. {\bf54}, 529 (2016).
\item  T. Camarillo, V. Mathur, T. Mitchell, and B. Ratra,
Publ. Astron. Soc. Pacif. {\bf130}, 024101 (2018).
\item X. Chen, S. Wang, L. Deng, and R. de Grijs, Astrophys. J. {\bf859}, 137 (2018).
\item A. K. Dambis, L. N. Berdnikov, Yu. N. Efremov,
A. Yu. Kniazev, A. S. Rastorguev, E. V. Glushkova,
V. V. Kravtsov, D. G. Turner, et al., Astron. Lett. {\bf41},
489 (2015).
\item G. S. Fishman and L. R. Moore, SIAM J. Sci. Stat.
Comput. {\bf7}, 24 (1986).
\item R. de Grijs and G. Bono, Astrophys. J. Suppl. Ser.
{\bf227}, 5 (2016).
\item E. Griv, I.-G. Jiang, and L.-G. Hou, Astrophys. J.
{\bf844}, 118 (2017).
\item A. I. Kobzar', Applied Mathematical Statistics. For
Engineers and Scientists (Fizmatlit, Moscow, 2006)
[in Russian].
\item V. Krishnan, S. P. Ellingsen, M. J. Reid, H. E. Bignall, J. McCallum, C. J. Phillips, C. Reynolds, and
J. Stevens, Mon. Not. R. Astron. Soc. {\bf465}, 1095
(2017).
\item D. Majaess, I. Dekany, G. Hajdu, D. Minniti,
D. Turner, and W. Gieren, Astrophys. Space Sci. {\bf363},
127 (2018).
\item F. Mignard, Astron. Astrophys. {\bf354}, 522 (2000).
\item I. I. Nikiforov, Astron. Rep. {\bf43}, 345 (1999).
\item I. I. Nikiforov, ASP Conf. Ser. {\bf209}, 403 (2000).
\item I. I. Nikiforov, ASP Conf. Ser. {\bf316}, 199 (2004).
\item I. I. Nikiforov and E. E. Kazakevich, Izv. GAO {\bf219} (4),
245 (2009).
\item I. I. Nikiforov and O. V. Smirnova, Astron. Nachr.
{\bf334}, 749 (2013).
\item I. I. Nikiforov and A. V. Veselova, Baltic Astron. {\bf24},
387 (2015).
\item I. I. Nikiforov and A. V. Veselova, Astron. Lett. {\bf44}, 81
(2018).
\item M. E. Popova and A. V. Loktin, Astron. Lett. {\bf31}, 171
(2005).
\item W. H. Press, S. A. Teukolsky, W. T. Vetterling, and
B. P. Flannery, Numerical Recipes in C (Cambridge
Univ. Press, Cambridge, UK, 1997).
\item A. S. Rastorguev, N. D. Utkin, M. V. Zabolotskikh,
A. K. Dambis, A. T. Bajkova, and V. V. Bobylev,
Astrophys. Bull. {\bf72}, 122 (2017).
\item M. J. Reid, private commun. (2014).
\item M. J. Reid, K. M. Menten, A. Brunthaler,
X. W. Zheng, T. M. Dame, Y. Xu, Y. Wu, B. Zhang,
et al., Astrophys. J. {\bf783}, 130 (2014).
\end{enumerate}


\end{document}